\documentclass{aa}
\newcommand{\EQ}{\begin{equation}}
\newcommand{\EN}{\end{equation}}
\newcommand{\EQA}{\begin{eqnarray}}
\newcommand{\ENA}{\end{eqnarray}}

\newcommand{\Eq}[1]{Eq.~(\ref{#1})}
\newcommand{\Eqs}[2]{Eqs~(\ref{#1}) and~(\ref{#2})}

\newcommand{\App}[1]{Appendix~\ref{#1}}
\newcommand{\Sec}[1]{Sect.~\ref{#1}}
\newcommand{\Fig}[1]{Fig.~\ref{#1}}
\newcommand{\FFig}[1]{Figure~\ref{#1}}
\newcommand{\Tab}[1]{Table~\ref{#1}}
\newcommand{\Figs}[2]{Figs~\ref{#1} and \ref{#2}}
\newcommand{\Tabs}[2]{Tables~\ref{#1} and \ref{#2}}
\newcommand{\bra}[1]{\langle #1\rangle}

{}
{}
{}
{}
{}
{}
{}
{}
{}
\newcommand{\yyy}{\hat{\mbox{\boldmath $y$}} {}}
\newcommand{\zz}{\hat{\mbox{\boldmath $z$}} {}}

\newcommand{\uu}{\mbox{\boldmath $u$} {}}

\newcommand{\BB}{\mbox{\boldmath $B$} {}}

\newcommand{\AAA}{\vec{A}}

\newcommand{\JJ}{\mbox{\boldmath $J$} {}}

\newcommand{\ff}{\mbox{\boldmath $f$} {}}

\newcommand{\EE}{\mbox{\boldmath $E$} {}}

\newcommand{\grav}{\mbox{\boldmath $g$} {}}
\newcommand{\nab}{\mbox{\boldmath $\nabla$} {}}

\newcommand{\curl}{{\rm curl} \, {}}

\newcommand{\DD}{{\rm D} \, {}}
\newcommand{\dd}{{\rm d} {}}

\newcommand{\ea}{{\rm et al. }}

\def\half{{\textstyle{1\over2}}}

\def\onethird{{\textstyle{1\over3}}}

\def\quarter{{\textstyle{1\over4}}}
\newcommand{\yapj}[3]{, #1, {ApJ }{#2}, #3}
\newcommand{\yapjl}[3]{, #1, {ApJ }{#2}, #3}

\newcommand{\yana}[3]{, #1, {A\&A }{#2}, #3}

\newcommand{\ygafd}[3]{, #1, {Geophys. Astrophys. Fluid Dyn. }{#2}, #3}
\newcommand{\yjfm}[3]{, #1, {JFM }{#2}, #3}

\newcommand{\ypp}[3]{, #1, {Phys. Plasmas }{#2}, #3}

\newcommand{\yprl}[3]{, #1, {Phys. Rev. Lett. }{#2}, #3}

\newcommand{\ynat}[3]{, #1, {Nat }{#2}, #3}

\newcommand{\ysph}[3]{, #1, {Solar Phys. } {#2}, #3}
\newcommand{\ypr}[3]{, #1, {Phys. Rev. } {#2}, #3}

\newcommand{\ybook}[3]{, #1, {#2} (#3)}
\newcommand{\yproc}[5]{, #1, in {#3}, ed. #4 (#5), p.#2}

\newcommand{\pgafd}[1]{, #1, {Geophys. Astrophys. Fluid Dyn. } (in press)}

\newcommand{\pmn}[1]{, #1, {MNRAS } (in press)}

\newcommand{\pana}[1]{, #1, {A\&A } (in press)}

\input{epsf}
\begin{document}
\title{Search for non-helical disc dynamos in simulations}
\author{Rainer Arlt\inst{1} and Axel Brandenburg\inst{2,3}}
\institute{
Astrophysikalisches Institut Potsdam, An der Sternwarte 16, D-14482 Potsdam, Germany
\and NORDITA, Blegdamsvej 17, DK-2100 Copenhagen \O, Denmark
\and Department of Mathematics, University of Newcastle upon Tyne, NE1 7RU, UK}

\date{\today}

\abstract{
The possibility of non-helical large scale dynamo action is investigated
using three-dimensional simulations of global accretion discs as well as
idealized local simulations without rotation and only shear. Particular
emphasis is placed on a certain correlation between vorticity and
azimuthal velocity gradient which has been predicted
to drive large scale dynamo action, independent of the presence or absence
of kinetic helicity.
In the global disc simulations two types of behaviours are found: those
which do show this type of velocity correlation and those which do not.
The former ones are typically also the cases where the resistivity is larger.
The latter ones show signs typical of dynamo action based on the usual helicity
effect. In the idealized simulations without rotation and just shear
the above correlation is found to be
particularly strong. In both cases there is, as expected, a systematic flux of
magnetic helicity through the midplane. However, very little helicity flux
leaves the domain through the top and bottom boundaries.
The idealized simulations reveal that much of this systematic flux
comes from the rotational component of the helicity flux and does not
contribute to its divergence.
\keywords{MHD -- Turbulence}
}

\maketitle

\section{Introduction}

A number of different simulations have now established the possibility
to generating strong {\it large scale} magnetic fields from turbulent motions
(Glatzmaier \& Roberts 1995, Brandenburg \ea 1995, Ziegler \& R\"udiger 2000,
Brandenburg 2001, hereafter referred to as B2001).
Although most of these simulations were devised to
explain magnetism in real astrophysical bodies, it remains debatable
whether or not the mechanisms that are at work in those simulations are
also those that are responsible for the generation of large scale fields
in astrophysical bodies. Up to modestly high
magnetic Reynolds numbers, currently accessible to simulations, the by far strongest
large scale dynamo effect is based on kinetic helicity of the flow.
This effect is described in standard text books (e.g.\ Moffatt 1978,
Krause \& R\"adler 1980), and is closely related to the inverse cascade
of magnetic helicity (Frisch \ea 1975, Pouquet \ea 1976). The
reason why one may suspect problems with such mechanisms in astrophysical
settings is that they tend to produce large scale magnetic fields that
are helical and that, owing to magnetic helicity conservation, such
helical fields can only be built up {\it slowly} on a resistive time scale
(B2001). Of course,
shear (or differential rotation) contributes strongly to the dynamo
and enhances the growth rate and final field strength, but this reduces the
resistively limited saturation
time of the dynamo only by a factor of 10--100 for the Sun
(Brandenburg \ea 2001, hereafter referred to as BBS2001),
or perhaps somewhat more for accretion discs. Typical growth times
will still be of the order of $10^6\,{\rm yr}$. Open boundaries
also tend to reduce the time scale, but this is typically at the expense of lowering the
final field strength (Brandenburg \& Dobler 2001, hereafter referred
to as BD2001).

The `helicity problem' was originally identified in attempts to
understand the `catastrophic' magnetic feedback on turbulent transport
coefficients such as turbulent diffusivity (Cattaneo \& Vainshtein 1991)
and the alpha-effect (Vainshtein \& Cattaneo 1992,
Cattaneo \& Hughes 1996). Although the original
arguments did not invoke magnetic helicity, subsequent work by
Bhattacharjee \& Yuan (1995) and Gruzinov \& Diamond (1995) related
the quenching to helicity conservation. Furthermore,
models using the proposed
quenching formulae reproduce the field evolution in the simulations
and those predicted by magnetic helicity conservation extremely well
(B2001, Fig.~21). Blackman \& Field (2000) pointed out that
catastrophic quenching of the $\alpha$-effect is peculiar to
flows in periodic domains where there is no loss of magnetic helicity.

A possible way out of the helicity problem is to produce large scale fields
without invoking kinetic helicity of the flow. That helicity is not
crucial for large scale dynamo action was already known since the
work of Gilbert \ea (1988), who found that flows that only lack parity invariance are
already capable of producing large scale dynamo action via an $\alpha$-effect.
The problem has
been investigated further in a recent paper by Zheligovsky \ea (2001),
who found that even parity invariant flows are capable of large-scale
dynamo action. In that case the dynamo works not via an $\alpha$-effect,
but through a negative turbulent magnetic diffusivity effect.
A related issue was brought up by Vishniac \& Cho (2001,
hereafter referred to as VC2001)
in an attempt to produce large scale non-helical dynamo action
that would survive in the large magnetic Reynolds number limit.
Their mechanism requires the presence of a certain correlation between
the azimuthal component of the vorticity and the azimuthal gradient of
the vertical velocity. If that is the case they predict the presence of
a strong vertical magnetic helicity current upwards. This could then drive a
field-aligned electromotive force that is proportional to the divergence
of this magnetic helicity current. This form of the electromotive force
would conserve magnetic helicity and was first proposed by Bhattacharjee
\& Hameiri (1986). 

The purpose of the present paper is to assess the viability and properties
of the mechanism proposed by VC2001 using numerical
simulations. The required correlation between the azimuthal derivative of
the vertical velocity and the azimuthal component of the vorticity
is expected to occur in the presence of shear.
It should thus be especially important in accretion discs.
We therefore begin by determining the presence of such a correlation
using global simulations of accretion discs (\Sec{S2}). However, in order to isolate
the proposed effect from the ordinary helicity effect, which is always
present because of rotation, we have also carried out some
idealized simulations of forced turbulence with shear,
but no rotation. The latter is expected to promote the correlation
anticipated by VC2001, but would not lead to kinetic helicity
in the flow. Of course,
real systems do rotate and have therefore also kinetic helicity. However,
at small and modestly large magnetic Reynolds numbers the dynamo effect
based on kinetic helicity is so much more powerful than other mechanisms
that it is necessary to suppress it artificially if one wants to study it in
isolation. This will be done in \Sec{S3}. For both the global disc
simulations as well as the idealized model we also determine
the resulting helicity fluxes, which turn out to be small and fluctuating about
zero, however. We conclude in \Sec{Sconcl} with summarizing remarks and
speculations concerning the viability of conventional helicity-driven
dynamos.

\section{A global disc simulation}
\label{S2}

\subsection{Description of the model}

In order to study the possibility of dynamo action of the form envisaged
by VC2001 we use the recent global hydromagnetic accretion disc simulation
by Arlt \& R\"udiger (2001). Unlike earlier global MHD disc simulations by
Armitage (1998) and Hawley (2000), the radial extent does here not cover the
entire disc, so inflow and outflow boundary conditions have to be applied
in the radial direction. A possible advantage of this is that a larger spatial
resolution per unit length is now possible. Like Armitage (1998), Arlt \& R\"udiger 
(2001) also use the ZEUS-3D code which, in turn, is
very similar to the code used by Hawley (2000). A version of ZEUS-3D that is
close to the one used by Arlt \& R\"udiger is described by Stone \& Norman (1992a,b) and 
Stone et al.\ (1992). The equations that are being solved are
\begin{eqnarray}
\frac{\partial\rho}{\partial t}+\nabla\cdot(\rho {\vec u})&=&0,\\
\frac{\partial{\rho \vec u}}{\partial t}+\nabla\cdot(\rho{\vec u}{\vec u})
&=&-\nabla p-\rho \nabla\Phi+{\vec J}\times {\vec B}+...,\label{dudt_zeus}\\
\frac{\partial {\vec B}}{\partial t}&=&\nabla\times({\vec u} \times {\vec B})
+\eta\nabla^2\vec B,
\end{eqnarray}
where $\rho$, ${\vec u}$, and ${\vec B}$ are density,
velocity, and magnetic field, respectively; $p$ is the pressure,
$\Phi$ is the gravitational potential (solely from a central
mass $M_\ast$), ${\vec J}=\vec{\nabla}\times{\vec B}/\mu_0$ is the
current density, $\mu_0$ is the vacuum permeability,
and $\eta$ is a constant magnetic diffusivity as implemented by 
D.~Elstner, Potsdam. The dots on the right hand side of \Eq{dudt_zeus}
indicate the presence of numerical viscosity for removing energy
at small scales. No energy equation appears as we deal with
an isothermal model where $p=c_{\rm s}^2\rho$. The sound speed,
$c_{\rm s}$, is 7\% of the Keplerian orbital velocity in the middle
of the simulated ring. Cylindrical polar coordinates, $(r,\phi,z)$, are
used and the entire azimuthal range from $\phi=0$ to $2\pi$
is considered. The model covers vertically the range from $z=-1$ to $+1$
which corresponds to 1.5--3 pressure scale heights, depending on the value of $r$. 

In most of the models we apply closed boundary conditions for the
flow on the $z$-boundaries, with the magnetic field penetrating
the boundaries at right angles. One of the models (Model~IX; see below) uses
open boundaries in the $z$-direction. The inner radial boundary is
open in all models; the mass flux is monitored on the inner boundary
and fed back into the domain on the
outer radial boundary with either a homogeneous or a Gaussian
infall pattern for the density. 
The maximum infall velocity is constant in time and over $z$
and is either $10^{-2}$ or $10^{-3}$ of the sound speed. The 
$\phi$-direction has periodic boundaries.

The initial configuration contains a relaxed disk with
a slow outflow on the inner boundary due to numerical viscosity. The 
density scale height varies between $H_\rho=0.33$ and
$0.66$ between the radii $r=4$ and $6$. In the absence of magnetic fields
the system is hydrodynamically stable; this
was verified numerically for up to 30~orbits
without producing any visible changes at the end of the simulation. The magnetic
field imposed to this configuration is merely a vertical $B_z$ field
with zero net flux through the vertical surfaces.

Table~\ref{tab1} summarizes the global runs used for this 
ana\-ly\-sis. The same numbering of the models as in 
Arlt \& R\"udiger (2001) is used. Two of the simulations are new on 
this list: Model~Va is a repetition of the configuration of 
Model~V, but with an initial magnetic field of mixed 
parity. The parity of the $B_z$ field was zero at the beginning
as was the parity of the emerging $B_\phi$ field. All other models 
start with an initial field parity of $-1$ (antisymmetry). Model~IX
is similar to Model~V, but uses outflow boundary conditions
for the vertical direction instead of closed boundaries.
In Table~\ref{tab1} the magnetic diffusivity $\eta$ is also given
for all runs. Since the considerations presented here were
made after the actual production runs had been performed, only
a limited number of fully three-dimensional snapshots were
available, which is the reason for a relatively coarse sampling.

\begin{table*}
\caption{\label{tab1}Global simulations used for the analysis.
The duration $t_{\rm end}$ of each run is given in orbits.
The infall velocities are given in units of the sound speed $c_{\rm s}$.
Symmetries refer to the type of initial $B_z$ fields. The kinetic energy
is based on the poloidal velocities only and is an average over
the last two orbits.}
\begin{small}
\begin{tabular}{lllccccc}
\hline
Run&Grid ($z$, $r$, $\phi$)&Radial boundary condition&$r$-range&$T_{\rm
orb}$ & $t_{\rm end}$ &Init.\ parity & $\eta$ \\
\hline
II& $31\times61\times351$ &homogeneous accretion $u_{\rm in}=-0.001c_{\rm s}$& 4.0--6.0&0.159 & 14.7 &antisym.&0.001 \\
III & $31\times61\times351$ &homogeneous accretion $u_{\rm in}=-0.01c_{\rm s}$& 4.0--6.0 &0.159 & \phantom{0}9.7&antisym.&0.001 \\
VIII& $31\times61\times351$ &homogeneous accretion $u_{\rm in}=-0.001c_{\rm s}$& 3.0--7.0 & 0.103 &  22.4&antisym.& 0.001 \\
V& $31\times61\times351$ &Gaussian accretion $u_{\rm in}=-0.001c_{\rm s}$&
 4.0--6.0& 0.159 & 18.4 &antisym.&0.01\phantom{0} \\
Va& $31\times61\times351$ &Gaussian accretion $u_{\rm in}=-0.001c_{\rm s}$&
 4.0--6.0& 0.159 & 12.4  &mixed&0.01\phantom{0}\\
VI& $31\times61\times351$ &Gaussian accretion $u_{\rm in}=-0.01c_{\rm s}$&
4.0--6.0&  0.159 & 16.1 &antisym.& 0.01\phantom{0} \\
IX& $31\times61\times351$ &Gaussian accretion $u_{\rm in}=-0.01c_{\rm s}$, $z$ open& 4.0--6.0&  0.159 & 16.9 &antisym.& 0.01\phantom{0} \\
\hline
\end{tabular}
\end{small}
\end{table*}

\subsection{The Vishniac--Cho correlation}

\begin{figure}\epsfbox{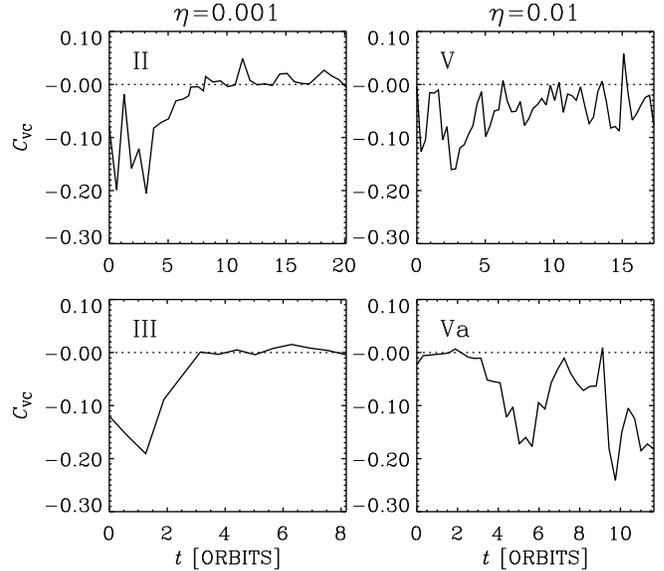}\caption[]{
Correlation of $\nabla_\phi u_z$ versus $\omega_\phi$
in global disc simulation; here of Models~II and III (left), as well as
Models V and Va (right).
}\label{Fvishcho_paper}\end{figure}

In the theory of VC2001 large scale magnetic field generation is possible
when there is a finite value of the correlation
\EQ
C_{\rm VC}\equiv\left\langle\omega_\phi\nabla_\phi u_z\right\rangle \Bigl/ 
  \sqrt{\langle\omega_\phi^2\rangle\langle(\nabla_\phi u_z)^2\rangle},
\EN
where $\nabla_\phi\equiv r^{-1}\partial_\phi$, 
and $\vec\omega=\curl\vec u$ is the vorticity. We begin by showing
$C_{\rm VC}$ as time series; see Fig.~\ref{Fvishcho_paper}.
The temporal averages of these time series (excluding the
initial five snapshots of each of the runs which correspond to 
a lapse of 2 to 4 orbits) are listed in Table~\ref{tab2}.
We found that the simulations can tentatively be 
divided into two groups: the first
group comprises the low-$\eta$ models showing decreasing
correlations (left column of \Fig{Fvishcho_paper}), the other group
has modestly large values of $\eta$ and is characterized by
a non-vanishing $C_{\rm VC}$ that is on the average about
5 to 10 times larger than in the low-$\eta$ cases
(right column of \Fig{Fvishcho_paper}).

\begin{figure}\epsfbox{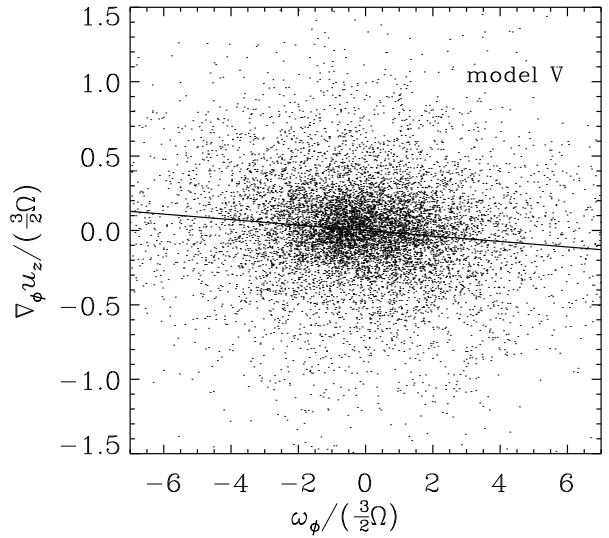}\caption[]{
Scatter plot of $\nabla_\phi u_z$ with $\omega_\phi$
for a snapshot of Model~V at $t=14.8\,T_{\rm orb}$. 
The straight line gives the least square fit, which has a
correlation coefficient of $C_{\rm VC}=-0.13$ in this example.
}\label{Fvishcho_scatter}\end{figure}

A scatter plot of $\nabla_\phi u_z$ versus $\omega_\phi$ is shown in
Fig.~\ref{Fvishcho_scatter} for the data of a snapshot from Model~V,
which has the largest (negative) correlation.
The plot contains points in the $(z,\phi)$ plane at $r=5$
which is in the middle of the computational domain.
The plot looks rather noisy, but one sees nevertheless a slight negative
correlation.

\epsfxsize=8.8cm

\epsfxsize=9.0cm

\subsection{Resulting magnetic field configurations}

\begin{figure}\epsfbox{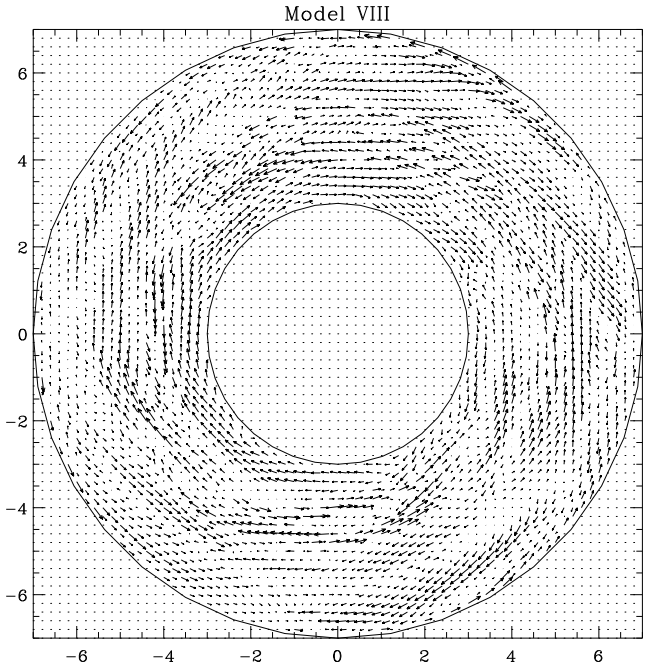}\caption[]{
Projection of magnetic field vectors in a horizontal slice at $z=-0.39$
of Model~VIII after 19.9 orbits.
}\label{Fshowdiskb_t90_20}\end{figure}

\begin{figure}\epsfbox{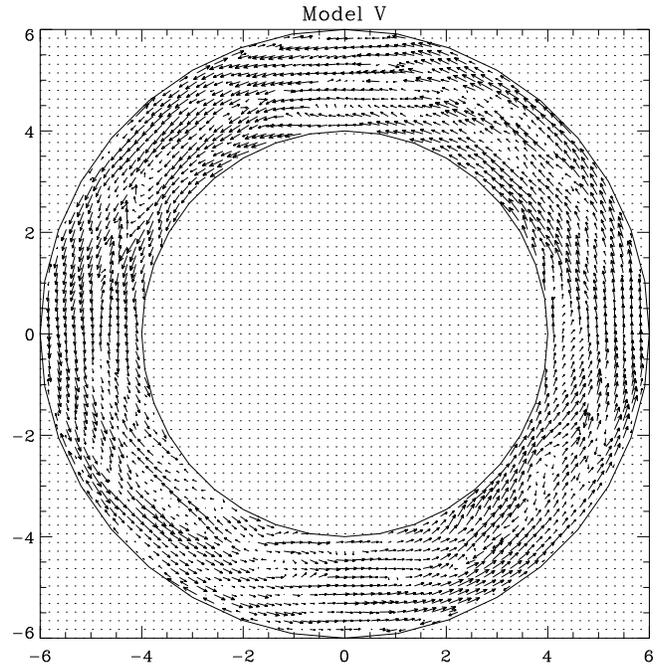}\caption[]{
Projection of magnetic field vectors in a horizontal slice at $z=-0.39$
of Model~V after 17.6 orbits.
}\label{Fshowdiskb_bgauss_65}\end{figure}

We first discuss the overall field structure.
Horizontal slices of the field at $z=-0.39$
are shown for Models~VIII (less resistive) and V (more resistive) in
Figs~\ref{Fshowdiskb_t90_20} and \ref{Fshowdiskb_bgauss_65}, respectively.
The former figure exhibits a spiral 
pattern whilst the latter is rather dominated
by intermediate scale structures or eddies, which is probably directly
a consequence of the larger magnetic diffusivity in that case.

\epsfxsize=8.8cm

\epsfxsize=8.8cm

\begin{figure}\epsfbox{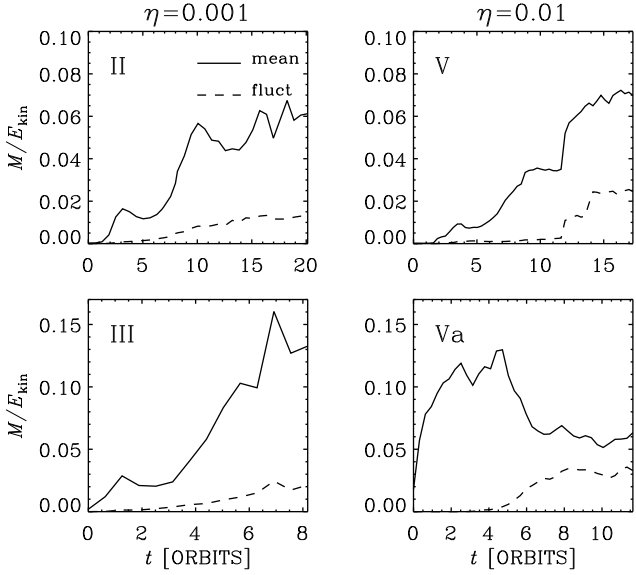}\caption[]{
Comparison of energies in the large scale
($\phi$-averaged) magnetic field (solid lines) and the
small-scale magnetic field (dotted lines) for Models~II and III (left),
as well as Models~V and Va (right). Note that the fluctuations are
larger in the cases where $\eta$ is larger.
}\label{Fhel_energy_paper}\end{figure}

Next we derive a number of averaged quantities from the simulations.
Throughout this section we denote azimuthal averages by an overbar, e.g.\
$\overline{\vec{B}}=\int\vec{B}\,\dd\phi/2\pi$.
In Fig.~\ref{Fhel_energy_paper} are shown 
the energies contained in the large scale field,
$M_{\rm mean}=\int\overline{\vec{B}}{}^2\dd V/(2\mu_0)$, and the
energies of the remaining fluctuations,
$M_{\rm fluct}=\int\vec{b}{}^2\dd V/(2\mu_0)$,
where $\vec{b}=\vec{B}-\overline{\vec{B}}$.
Like in
Fig.~\ref{Fvishcho_paper}, the temporal evolution behaviours separate
into the same two groups: the low-$\eta$ (less resistive) runs
which show significant energies in the
large scale field, and models with larger $\eta$ that are more resistive,
but better able to generate
fluctuation energies of at least 50\% of the large scale energy.
We note that there is one model (Model~IX, not shown) where at the end the
energy of magnetic fluctuations exceeds the large scale magnetic energy.
The fact that the energy of the mean field is typically larger than
that of the fluctuating field is somewhat surprising. A possible reason
could be that the memory of the initial mean field has not yet been lost.
It is also possible, however, that it is because of the global geometry
and the shear that a strong large scale field is more easily established
when the magnetic Reynolds number is large.

\epsfxsize=8.8cm

\begin{figure}\epsfbox{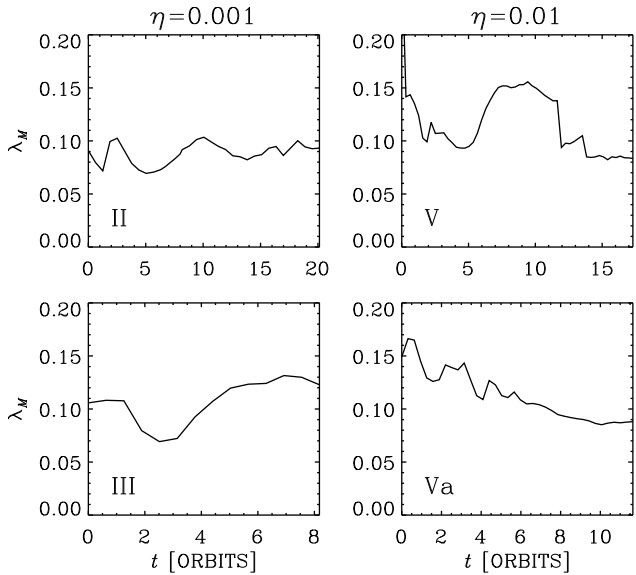}\caption[]{
Comparison of the magnetic Taylor microscale $\lambda_{\rm M}$ 
for Models~II and III (left), as well as V and Va (right).
}\label{Fhel_lambda_paper}\end{figure}

Another distinction between the two groups of runs is given
by the magnetic Taylor microscale, $\lambda_{\rm M}$, which we define
here via $\lambda_{\rm M}^2=\langle\vec{B}^2\rangle/\langle\vec{J}^2\rangle$.
In Fig.~\ref{Fhel_lambda_paper} we show the value of $\lambda_{\rm M}$
for the same four models as in Fig.~\ref{Fvishcho_paper}.
The quantity $2\pi\lambda_{\rm M}$ characterized the typical thickness
of flux structures. Its significance is that in runs with dynamo action
$\lambda_{\rm M}$ tends to increase with time until it reaches saturation
(e.g.\ Brandenburg \ea 1996). Conversely, when the field is amplified just
by field compression the value of $\lambda_{\rm M}$ decreases somewhat
with time. This is what happened in the more resistive runs ($\eta=0.01$).
These were actually the runs that did show evidence for a finite
Vishniac--Cho correlation. In contrast, the less resistive runs
($\eta=0.001$) do not show any such trend.

\begin{figure}\epsfbox{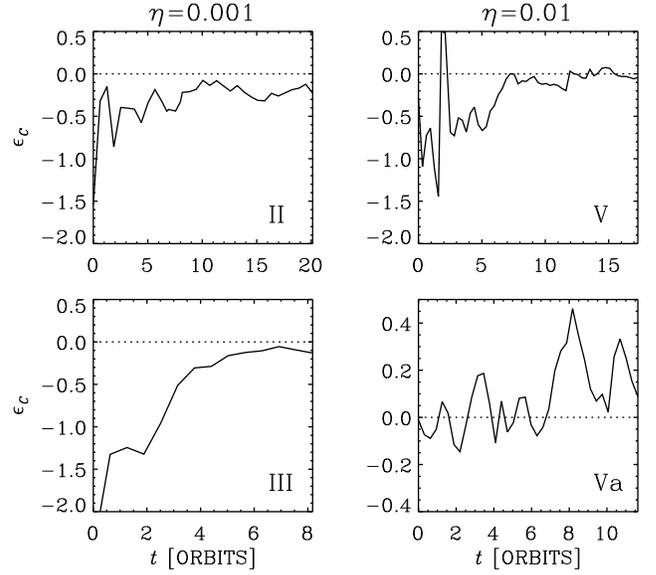}\caption[]{
Evolution of the nondimensional large-scale current helicity parameter
(upper disc plane) for Models~II and III (left),
as well as Models~V and Va (right).
}\label{Fhel_epsilon_paper}\end{figure}

A useful quantity for assessing the importance of helicity in the
large scale field is to look at the nondimensional quantity
\EQ
\varepsilon_C={C_{\rm mean}/k_1\over M_{\rm mean}},
\EN
where $C_{\rm mean}=\int\overline{\vec{J}}\cdot\overline{\vec{B}}\,\dd V$
is the current helicity of the mean field.
In BD2001 the value of $\varepsilon_C$ was found to be of order
unity (between 1--2) for the models with a halo.
In Fig.~\ref{Fhel_epsilon_paper} we show the value of $\varepsilon_C$
for the global accretion disc runs. The less resistive models
show negative current helicity whilst the more resistive ones
show vanishing or positive current helicity toward the end of the run.
In any case, the values of $|\varepsilon_C|$ are smaller compared
with the models studied in BD2001 where $|\varepsilon_C|={\cal O}(1)$.
This is not surprising, because
in BD2001 the kinetic helicity was close to the maximum value, which
would be an unrealistic assumption for any astrophysical body.

\epsfxsize=8.8cm

\epsfxsize=8.8cm

We now discuss the sign of $\varepsilon_C$. In BD2001 the sign of the
kinetic helicity was positive, and so was the sign of the {\it small
scale} current helicity. The sign of the large scale current helicity
is typically opposite, i.e.\ negative in that case. In the present case,
where we consider the upper disc plane, the kinetic helicity is negative,
so we would expect a positive value of $\varepsilon_C$, which is not what
we find (except for one of the more resistive cases, Model Va). However,
the unusual sign of $\overline{\vec{J}}\cdot\overline{\vec{B}}$ is
primarily related to an unusual sign of the $\alpha$-effect itself.
This is because in the steady state $\alpha$-effect and turbulent
diffusion balance each other, so the sign of $\alpha$ must coincide
with the sign of $\overline{\vec{J}}\cdot\overline{\vec{B}}$.
In discs, however, the sign of $\alpha$ is negative (in the upper disc plane),
so $\overline{\vec{J}}\cdot\overline{\vec{B}}$ must also be negative, and
hence $\varepsilon_C$ is negative, as observed. The perhaps most
convincing explanation for the negative $\alpha$ is that intense parts of
a flux tube contract (to maintain pressure balance along field lines),
but are also most
buoyant. If this contraction is stronger than the expansion associated
with the rise into a less dense medium, then $\alpha$ will be negative
(Brandenburg 1998, see also R\"udiger \& Pipin 2000).

We shall now return to the question of whether there is any evidence for the
presence of a dynamo effect as envisaged by VC2001. We therefore need to
look at the possibility of magnetic helicity fluxes through the domain.

\begin{table*}
\caption{\label{tab2}
Results after the magnetic field was switched on. All values are 
temporal averages excluding the first five snapshots (about 2~to
4~orbital periods).
The first group of models (II, III, and VIII) comprises less resistive
runs, whilst the second group (Models V, Va, VI, and IX) refers to the 
more resistive ones.}
\begin{tabular}{lccrrrrccc}
\hline
Run&$u_{\rm rms}$&$E_{\rm kin}$&
\multicolumn{2}{c}{$2\eta C_{\rm mean}/(\mu_0 u_{\rm rms}E_{\rm kin})$}&
\multicolumn{2}{c}{$2\eta C_{\rm fluc}/(\mu_0 u_{\rm rms} E_{\rm kin})$}&
$Q_{\rm mean}/$&slope&$C_{\rm VC}$\\
  & & &North&South&North&South &$(\mu_0 u_{\rm rms} E_{\rm kin})$ & &\\
\hline
II& 3.7 & $3.6\times 10^5$& $-7.8\cdot10^{-6}$& $+1.5\cdot10^{-6}$ & $-5.0\cdot10^{-6}$ & $+5.6\cdot10^{-6}$ & $+0.00006$ &
          $-0.0024$ & $-0.012$ \\
III& 7.5& $6.9\times 10^5$& $-4.0\cdot10^{-6}$& $+9.3\cdot10^{-7}$& $-2.9\cdot10^{-6}$ & $+2.3\cdot10^{-6}$ & $+0.00007$&
          $-0.0005$ & $+0.003$\\
VIII& 7.6 &$4.9\times 10^5$ &$-5.4\cdot10^{-7}$ & $+3.5\cdot10^{-7}$ & $-2.8\cdot10^{-6}$ & $+3.5\cdot10^{-6}$ &$+0.00029$ &
          $-0.0018$ & $-0.005$\\
\hline
V& 3.8 & $1.4\times 10^6$& $-2.2\cdot10^{-5}$& $+3.9\cdot10^{-5}$& $-2.0\cdot10^{-7}$   & $+1.1\cdot10^{-5}$ &$-0.00013$ &
        $-0.0086$ & $-0.050$ \\
Va& 5.3 &$3.9\times 10^5$& $+6.9\cdot10^{-5}$& $+6.6\cdot10^{-5}$& $+2.4\cdot10^{-4}$& $-2.6\cdot10^{-5}$ & $-0.00038$&
        $-0.0096$ & $-0.086$ \\
VI& 4.0 &$1.7\times 10^6$& $-5.0\cdot10^{-5}$& $+3.3\cdot10^{-5}$& $-5.0\cdot10^{-6}$  & $+2.1\cdot10^{-6}$  &$-0.00157$ &
        $-0.012\phantom{0}$  & $-0.060$ \\
IX& 2.9 &$2.9\times 10^5$& $-1.9\cdot10^{-4}$& $+1.6\cdot10^{-4}$& $-1.4\cdot10^{-4}$ & $+1.9\cdot10^{-4}$ &$+0.00007$ &
        $-0.0053$ & $-0.047$ \\
\hline
\end{tabular}
\end{table*}

\subsection{Magnetic helicity flux}

In VC2001 it was argued that, although the overall magnetic helicity is
small, there could still be a significant (spatially constant) flux
of magnetic helicity vertically through the domain. The numerical procedures
for evaluating gauge invariant magnetic helicity and magnetic helicity
flux in cylindrical geometry with open boundaries in the $r$ and $z$
directions are not yet available. However, for the present purpose most important
is the contribution from the large scales. If we adopt horizontal averages
(over $r$ and $\phi$), the mean fields are one-dimensional and the mean
magnetic vector potential can be obtained simply by integration. The
corresponding magnetic helicity and magnetic helicity fluxes of the
mean field can then be calculated quite easily (see the appendix of BD2001). In
Fig.~\ref{Fhel_q_paper2} we plot, for the four models, the magnetic helicity
flux, $Q_{\rm mean}=Q_{\rm mean}^{(2)}-Q_{\rm mean}^{(1)}$, 
out of the domain through the two
boundaries at $z=z_1$ and $z_2$. Here, $Q_{\rm mean}^{(1)}$ 
and $Q_{\rm mean}^{(2)}$ denote the upward helicity fluxes at $z=z_1$
and $z_2$, respectively; see \App{App1}.

\epsfxsize=8.8cm\begin{figure}\epsfbox{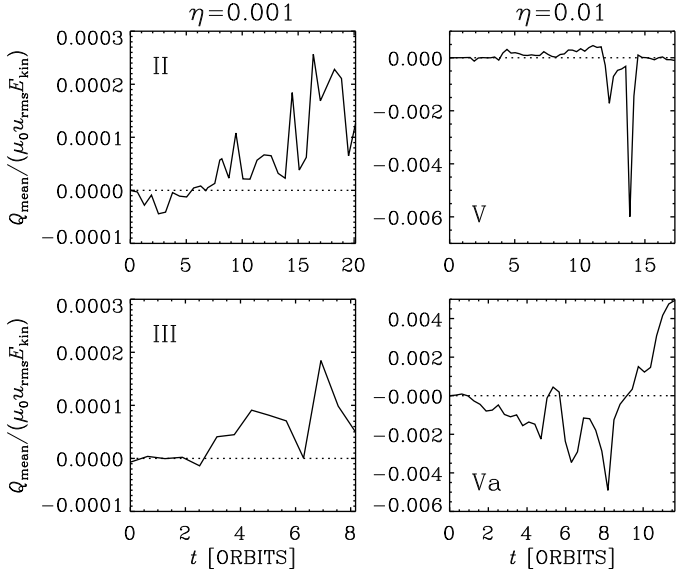}\caption[]{
Estimates of the magnetic helicity flux $Q_{\rm mean}$ out through the vertical
boundaries. The dotted lines give the zero value.
}\label{Fhel_q_paper2}\end{figure}

The mean outward flux, expressed in `dynamical' units, $\mu_0 u_{\rm rms}
E_{\rm kin}$, is small ($\sim10^{-4}$). A small outward flux was also
found in the case of helical turbulence (BD2001). Furthermore, 
the signs tend to be different in the cases where $\eta$ is 
small ($Q_{\rm mean}>0$ for $\eta=10^{-3}$) and where 
it is larger ($Q_{\rm mean}<0$ for $\eta=10^{-2}$). It is not clear
whether this result indicates any significant departure from zero,
because the magnetic helicity on the two sides about the midplane of
the disc are expected to be different. Thus, equal losses or gains on
the two surfaces ($z=z_1$ and $z_2$) should result in zero net magnetic
helicity flux. It is therefore now necessary to determine the mean 
{\it upward} fluxes of magnetic helicity on the two sides, 
$Q_{\rm mean}^{(2)}$ and $Q_{\rm mean}^{(1)}$. Its average is denoted by
$Q_{\rm mean}^{\rm(up)}=\half[Q_{\rm mean}^{(2)}+Q_{\rm mean}^{(1)}]$.
If there is indeed a systematic upwards flux through the two boundaries,
this quantity should be finite and positive. Instead, we find it to be small.
In order to check whether this is the result of some cancellation,
we need to consider the magnetic helicity fluxes in smaller sub-volumes.

A difficulty associated with calculating magnetic helicity and
magnetic helicity fluxes separately in two sub-volumes (e.g., above and
below the midplane) is that we want to make sure that the sum of the two is
equal to the total magnetic helicity calculated earlier. This will be the
case provided the magnetic helicity in each sub-domain is calculated
using the same gauge that also made the helicity of the full domain
gauge invariant. This then also allows one to calculate the integrated
magnetic helicity fluxes out of each sub-domain. The corresponding
formulae are given in \App{App2}.

It turns out that the helicity fluxes out of each sub-volume are
actually large but of opposite sign. This means that there is actually
a large magnetic helicity flux through the midplane, but not through
the upper and lower boundaries. Having fixed the gauge such that
$\int\overline{\vec{A}}\cdot\overline{\vec{B}}\,\dd z$ is equal to
the helicity of BD2001 for the {\it full} domain, we can also calculate
the local magnetic helicity fluxes. We denote these by
$Q_{\rm mean}^{\rm(mid)}$ (if evaluated at the midplane), or by
$Q_{\rm mean}^{(z)}$ (if calculated for all values of $z$).
(We recall that the volume integrated divergence of the helicity flux,
which enters on the right hand side of the magnetic helicity equation,
is given by the {\it difference} of $Q_{\rm mean}^{(z)}$ between 
the two $z$-boundaries and is shown in Fig.~\ref{Fhel_q_paper2} 
and Tab.~\ref{tab2}.)

\begin{figure}\epsfbox{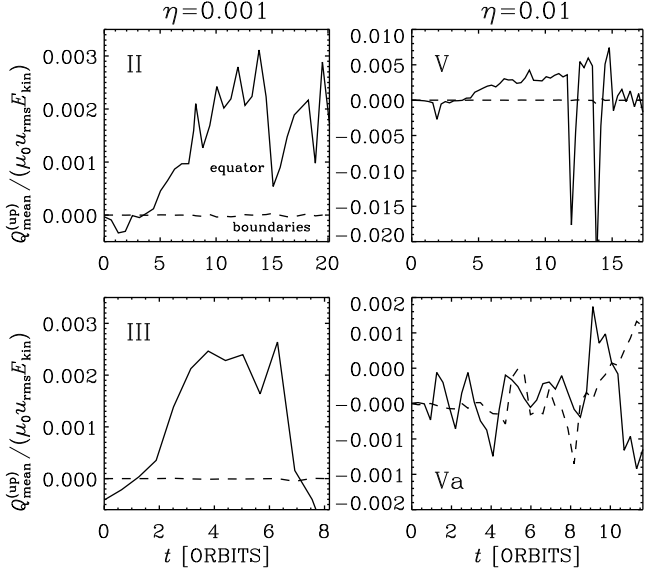}\caption[]{
Estimates of the integrated magnetic helicity fluxes
at the midplane, $Q_{\rm mean}^{\rm(mid)}$, (solid line) and on
the boundaries, $Q_{\rm mean}^{\rm (up)}$ (dotted line).
\label{Fhel_qupinew_paper}}
\end{figure}

In Fig.~\ref{Fhel_qupinew_paper} we plot $Q_{\rm mean}^{\rm(mid)}$ and
compare with the averaged boundary fluxes $Q_{\rm mean}^{\rm(up)}$.
It turns out that $Q_{\rm mean}^{\rm(mid)}$ is indeed mostly positive,
as predicted by VC2001, but this flux is not sustained all the way
to the boundaries: $Q_{\rm mean}^{\rm(up)}$ is virtually zero by
comparison. An exception is Run~Va, where $Q_{\rm mean}^{\rm(mid)}$
shows large variations about zero and $Q_{\rm mean}^{\rm(up)}$ begins
to deviate systematically from zero. (We recall that this is the run
where the initial field had mixed parity.)

\epsfxsize=8.8cm

\begin{figure}\epsfbox{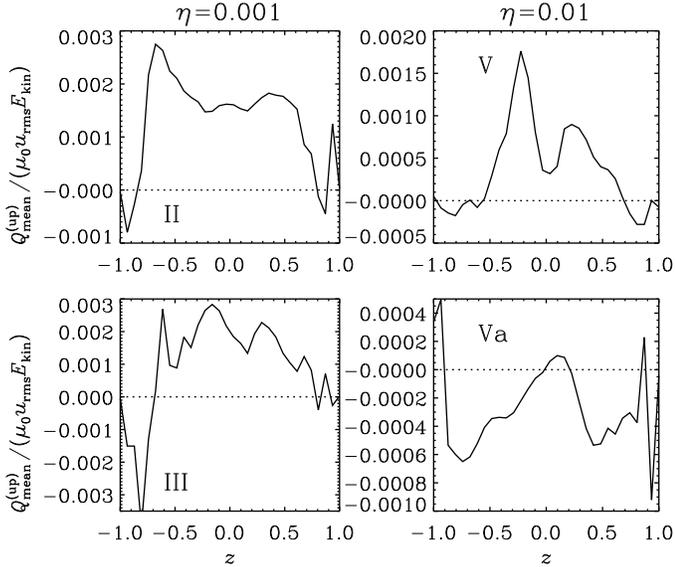}\caption[]{
Vertical distribution of the estimates of the upward magnetic 
helicity flux $Q_{\rm mean}^{\rm(up)}$ through planes at various 
locations $z$ (negative sign means downward). The values at each $z$
are averages over time, excluding the first five snapshots of the
simulations.
\label{Fhel_qupi_paper}}
\end{figure}

In order to see whether the magnetic helicity flux at the midplane
is typical of the entire interior of the simulation domain, we plot
in Fig.~\ref{Fhel_qupi_paper}
the vertical distribution of the magnetic helicity flux, which was derived
from horizontal averages of field and flow and then averaged in time
(again excluding the first 2 to 4~orbits).
Fig.~\ref{Fhel_qupi_paper} shows that a positive (i.e.\ upwards) flux
of magnetic helicity is indeed typical of the interior of the entire domain,
and that it vanishes only near the boundaries. Thus the boundaries seem to
play an important role, which may of course be unrealistic. We note,
however, even Model~IX with open boundaries does show a rapid drop of
magnetic helicity flux near the $z$-boundaries. Clearly, an
abrupt change of this flux implies
generation and destruction of magnetic helicity
near the vertical boundaries.

\epsfxsize=8.8cm

To summarize the global disc simulations, the correlation
anticipated by VC2001 is present, provided the magnetic
diffusivity is not too small. If the magnetic diffusivity is
smaller, the magnetic field tends to be stronger and can become more
important and may hence be able to suppress this correlation.
Nevertheless, these are also the cases which show most clearly
a systematic magnetic helicity flux within the simulation domain,
even though it is unable to leave or enter it through the boundaries.
It is difficult to say whether the effect of VC2001 was really responsible
for the field generation found in the disc simulations.
We recall that in the present simulations there is also some evidence for a
$\alpha$-effect, although it is based on a rather
noisy correlation between the turbulent electromotive force and the mean
field; see Arlt \& R\"udiger (2001).
In the following section we isolate the VC-effect by
studying a more idealized model with no net helicity. This model
also allows for longer runs and therefore clearer statistics.

\section{An idealized model with shear and stratification,
but no rotation nor helicity}
\label{S3}

In order to study the effect proposed by VC2001 in isolation we need to
suppress artificially the ordinary dynamo effect due to kinetic helicity
of the flow. Following VC2001 the crucial factor that is necessary for
causing the anticipated correlation is shear, but
not rotation. To check this we have used a simulation of externally
driven turbulence with zero net helicity.  This is done by using the
forcing function of B2001, but with helicity whose sign changes randomly.
In addition, the model
is supplemented by the effects of sinusoidal shear (as in BBS2001)
and vertical stratification with a conducting halo above a
turbulent layer (as in BD2001).

\subsection{Description of the model}

We solve the isothermal compressible MHD equations for the logarithmic
density $\ln\rho$, the velocity $\uu$, and the magnetic vector potential
$\AAA$,
\EQ
{\DD\ln\rho\over\DD t}=-\nab\cdot\uu,
\EN
\EQ
{\DD\uu\over\DD t}=-c_{\rm s}^2\nab\ln\rho+{\JJ\times\BB\over\rho}
+{\mu\over\rho}(\nabla^2\uu\!+\!\onethird\nab\nab\!\cdot\!\uu)+\ff,
\label{dudt}
\EN
\EQ
{\partial\AAA\over\partial t}=\uu\times\BB-\eta\mu_0\JJ-\nab\phi,
\label{dAdt}
\EN
where ${\rm D}/{\rm D}t=\partial/\partial t+\uu\cdot\nab$ is the
advective derivative, $\BB=\nab\times\AAA$ is the magnetic field, and
\EQ
\ff=\ff_{\rm turb}+\ff_{\rm shear}+\grav
\EN
is the sum of a random forcing function, $\ff_{\rm turb}$ (specified in B2001),
a sinusoidal shear profile, $\ff_{\rm shear}=S_0(\mu/\rho)\yyy\sin x$,
and a periodic gravity potential, $\grav=\half g_0\zz\sin(z/2)$.
In all calculations we assume the gauge $\phi=0$ (vanishing electrostatic
potential), but for some of the analysis we also adopt the gauge
$\nab\cdot\AAA=0$. Instead of the
dynamical viscosity $\mu$ ($=\mbox{const}$) we will in the following
refer to $\nu\equiv\mu/\rho_0$, where $\rho_0$ is the mean density in
the domain ($\rho_0=\mbox{const}$ owing to mass conservation).

We use nondimensional units where $c_{\rm s}=k_1=\rho_0=\mu_0=1$. Here,
$c_{\rm s}=\mbox{const}$ is the isothermal sound speed, $k_1$ the smallest wavenumber
of the two horizontal directions (so its size is $2\pi$ in the horizontal
direction). The vertical extent of the domain is $4\pi$. Periodic boundary
conditions are adopted in all three directions. The wavenumber of the
forcing is $k=k_{\rm f}=5$.
As in BBS2001 the forcing amplitude is $f_0=0.01$ and the nominal shear is
$S_0\equiv|\partial u_y/\partial x|_{\max}=1$.
This means that the resulting shear velocity (in the absence of magnetic stresses)
is then also of order unity, i.e.\ close to the speed of sound, and the
turbulent rms velocity is about a hundred times smaller.
As in BD2001, the gravitational potential varies sinusoidally in $z$ with an
amplitude $g_0=0.5$, so the density contrast is $\Delta\ln\rho\approx1$.
The main parameter that is varied in the models considered below is the
magnetic diffusivity $\eta$, which is in the range $(1...5)\times10^{-4}$.

\begin{table}[t!]\caption{
Summary of the main properties of various runs. The kinetic energy,
$E_{\rm kin}$, is based on the poloidal flow only, whilst $E_{\rm kin}^{\rm(tot)}$
refers to the total kinetic energy (including the shear motion).
In Runs~A--C the resolution is $60^2\times120$ meshpoints and in Run~C2
it is $120^2\times240$ meshpoints. The magnetic energies in Run~C2 are
only lower limits, because the field has not reached final saturation yet.
}\vspace{12pt}\centerline{\begin{tabular}{ccccccc}
Run &$\nu$ &  $\eta$   & $E_{\rm kin}$& $E_{\rm kin}^{\rm(tot)}$
& $M_{\rm fluct}$ & $M_{\rm mean}$ \\
\hline
A & 0.01 &$\!\!5\times10^{-4}$&  0.0030 & $2.2$  & 0.005 & 0.002 \\
B & 0.01 &$\!\!2\times10^{-4}$&  0.0030 & $0.7$  & 0.018 & 0.002 \\
C & 0.01 &    $10^{-4}$   &  0.0035 & $0.23$ & 0.013 & 0.001 \\
C2& 0.005&    $10^{-4}$   &  0.005  & $0.35$ &$\!\!>0.012$&$\!\!>0.0004$\\
\label{T1}\end{tabular}}\end{table}

Once the magnetic field becomes strong the shear motion becomes reduced
significantly due to magnetic forces. We define the total kinetic energy
(per unit surface) as $E_{\rm kin}^{\rm(tot)}=\half\int_{z_1}^{z_2}
{\overline{\rho\vec{u}^2}}\dd z$, where $z_1=-\pi$ and $z_2=\pi$ are
the boundaries between halo and disc plane. (This expression for
$E_{\rm kin}^{\rm(tot)}$ includes the energy contained in
the shear, in contrast to $E_{\rm kin}$ which does not.)
When the field is still weak we have
$E_{\rm kin}^{\rm(tot)}\approx2.4$, but once the field is strong
this value gets significantly reduced by magnetic stresses; see \Tab{T1}. One should
keep this in mind when comparing the magnetic energies for the
different runs.
In the following we use rms values of velocity and magnetic field for
normalization purposes; these quantities are defined in terms of
$E_{\rm kin}$ and $M$ via $u_{\rm rms}=\sqrt{2E_{\rm kin}/(\bra{\rho}L_z)}$
and $B_{\rm rms}=\sqrt{2M/L_z}$, respectively.

\subsection{The Vishniac--Cho correlation}

\begin{figure}[t!]\epsfbox{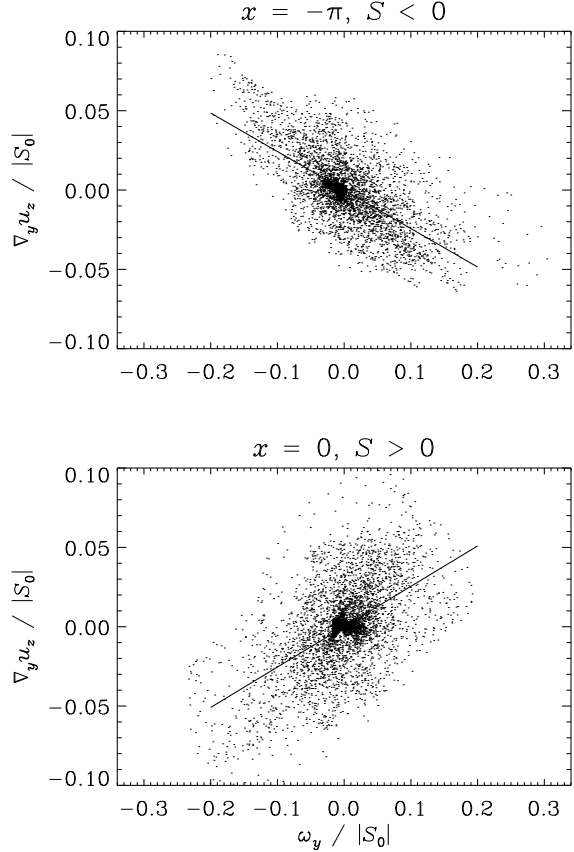}\caption[]{
Scatter plot showing the correlation between $\nabla_y u_z$ and
$\omega_y$ for a snapshot taken at a late time of the run with
$\eta=5\times10^{-4}$. In this example the correlation coefficients are
$-0.66$ (for $x=-\pi$, where $S<0$) and $+0.55$ (for $x=0$, where $S>0$).
}\label{Fpethan}\end{figure}

In \Fig{Fpethan} we plot the resulting correlation between $\nabla_y u_z$
($\equiv\partial u_z/\partial y$) and $\omega_y$ for a particular snapshot
at the end of Run~A. The anticipated effect is relatively
well pronounced -- much more than in the global disc simulations.
As expected (see VC2001), its sign changes where the local shear,
$S=\partial u_y/\partial x$, is reversed.
This supports the validity of the basic result of VC2001 that
such a correlation exists owing to the presence of shear.

It turns out that the Vishniac--Cho correlation is in fact the most
significant correlation coefficient that changes sign when shear changes sign. In \App{App3}
we have calculated, for this flow, all 81 correlation coefficients of
$\bra{u_{i,j}u_{k,l}}$. The Vishniac--Cho correlation is given by the
sum of $\bra{u_{x,z}u_{z,y}}$ and $-\bra{u_{z,x}u_{z,y}}$. These two
terms are indeed the most important ones. We note that the correlators
$\bra{u_{i,j}u_{j,i}}$ (for $i\neq j$) are also relatively large, but they do not
change sign when the sign of the shear changes.

Next, we need to check whether this flow
is capable of dynamo action and whether large scale fields can
be generated. If so, then this effect should be associated with
a significant vertical transport of magnetic helicity (VC2001).
Whether or not such a flux really helps the dynamo needs of course
to be seen.

\epsfxsize=8.2cm

\subsection{Dynamo action}

\begin{figure}[t!]\epsfbox{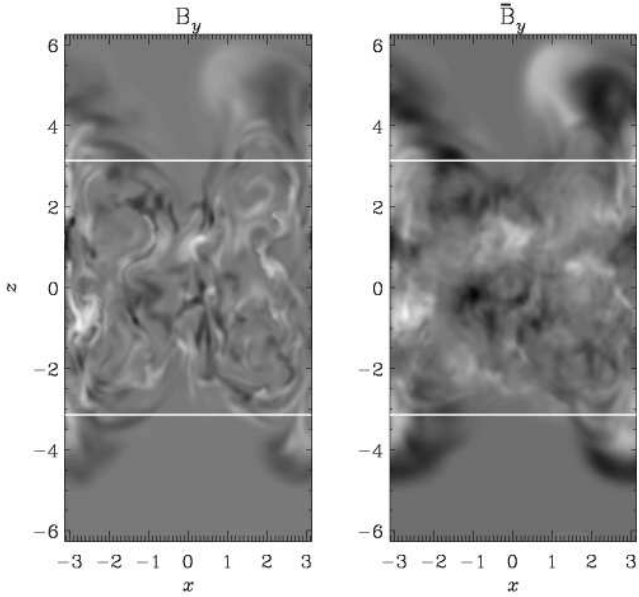}\caption[]{
Vertical slice of $B_y$ (left) and its $y$-average (right) for Run~C2.
The boundaries between halo
and disc plane are indicated by white lines. The resolution of this
run is $120^2\times240$ meshpoints and $t=810$, corresponding to
$\lambda t=4.3$. Dark and light
shades indicate negative and positive values, respectively.
}\label{Fpslice_mhdp}\end{figure}

\begin{figure}[t!]\epsfbox{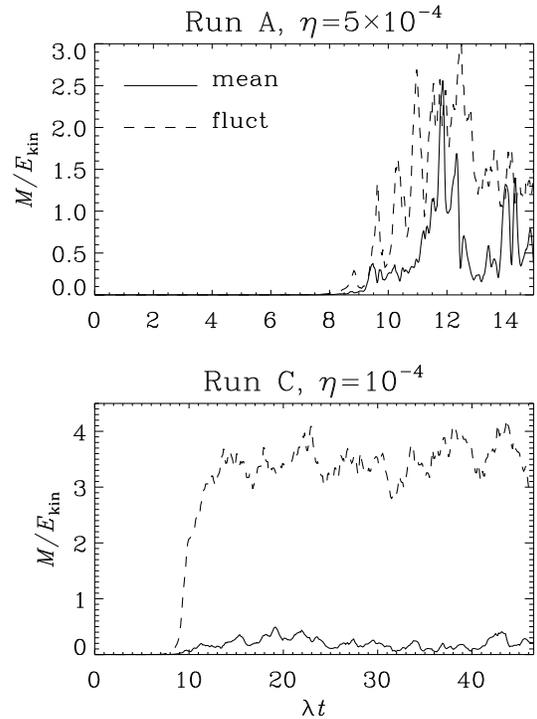}\caption[]{
Normalized magnetic energies in the mean and fluctuating components for Runs~A
and C. Time is given in units of the inverse growth
rate ($\lambda=0.0012$ for Run~A and $\lambda=0.0053$ for Run~C).
}\label{Fphelm_comp}\end{figure}

In this model we find dynamo action provided the magnetic Reynolds
number is large enough. In \Tab{T1} we give various parameters for
the models considered. In Run~A where $\eta=5\times10^{-4}$ (the
value used in BBS2001) the dynamo growth rate,
$\lambda\equiv\dd\ln B_{\rm rms}/\dd t$, is about $\lambda=0.0012$.
This is much less than in the case with helical forcing where
$\lambda=0.015$.
As in B2001 and BBS2001 the initial field strengths are about $10^{-6}$.
Thus, in the present case the saturation time is about
$\lambda^{-1}\ln10^6\approx10^4$, compared to about $10^3$ in BBS2001.
For $\eta=10^{-4}$ (Run~C) the growth rate is about five times larger (0.0053)
and the resulting saturation time correspondingly shorter. The field
is concentrated at small scales and shows some loop-like pattern
within the turbulent layer; see \Fig{Fpslice_mhdp}.
In \Fig{Fphelm_comp} we plot the evolution of magnetic
energies of the mean and fluctuating field components
within the turbulent zone, $|z|\le\pi$.
Here, mean fields are defined with respect to averaging in the toroidal
direction. Large scale magnetic fields are not present in the
bulk of the turbulent layer, but can be seen in the halo, especially
at later times.

\epsfxsize=8.6cm

In Run~A where $\eta=5\times10^{-4}$ there is significant magnetic
energy in the mean magnetic field (upper panel of \Fig{Fphelm_comp}),
but this mean field diminishes and the fluctuating field gains in
strength as the magnetic diffusivity is lowered to $\eta=10^{-4}$ (Run~C).

\epsfxsize=8.2cm

Next we look at the evolution of magnetic and current helicities;
\EQ
H=\int_{z_1}^{z_2}\overline{\vec{A}\cdot\vec{B}}\,\dd z,\quad
C=\int_{z_1}^{z_2}\overline{\vec{J}\cdot\vec{B}}\,\dd z,
\EN
where $z_1$ and $z_2$ are the boundaries of the turbulent subdomain.
In the present case, $z_1=-\pi$ and $z_2=+\pi$, as indicated by white
lines in \Fig{Fpslice_mhdp}.
Overbars denote horizontal averaging, so the helicities are really volume
integrals, but they are normalized by the horizontal surface area.
The definition of $H$ depends on the gauge, but by extrapolating the
field onto a periodic domain using potential fields one can define a
gauge independent magnetic helicity (Berger \& Field 1984). This is
also the one used in this paper. The gauge independent magnetic
helicity of the mean-field is particularly important and requires
special care due to the fact that our vector and gauge potentials
are periodic in the horizontal directions; see BD2001\footnote{We
use this opportunity to point out a sign error in their Eq.~(9),
where it should read $\vec{A}_0=\overline{\vec{A}}_0
-\vec{\nabla}_\perp\times(\psi\vec{\hat{z}})$. The contributions
from the mean field remain however unaffected.} for details.

\begin{figure}[t!]\epsfbox{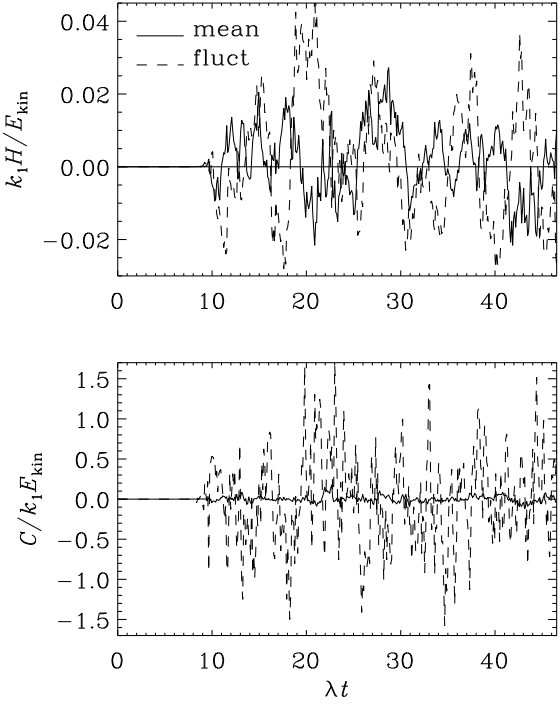}\caption[]{
Magnetic and current helicities in the mean and fluctuating components
for the run with $\eta=10^{-4}$. $H$ and $C$ are given in units of
$\mu_0\rho_0 u_{\rm rms}^2 L_z/k_1$ and $\rho_0 u_{\rm rms}^2 L_z k_1^2$,
respectively.
}\label{Fphelh}\end{figure}

As expected, because there is no net helicity in the forcing of the flow,
there are also no net magnetic and current helicities in the fields;
see \Fig{Fphelh}. The fluctuations of magnetic helicity are generally
weak, but the contributions from the mean and fluctuating fields are
comparable. For the current helicity there is a clear dominance of the
small scale fields over the large scale fields. This is explained by
the fact that the current helicity has two $k$-factors more than the
magnetic helicity and hence the ratio of small scale to large scale
contributions are larger by a factor $(k_{\rm f}/k_1)^2$ for current
helicity relative to magnetic helicity.

\epsfxsize=8.2cm

\epsfxsize=8.4cm\begin{figure}[t!]\epsfbox{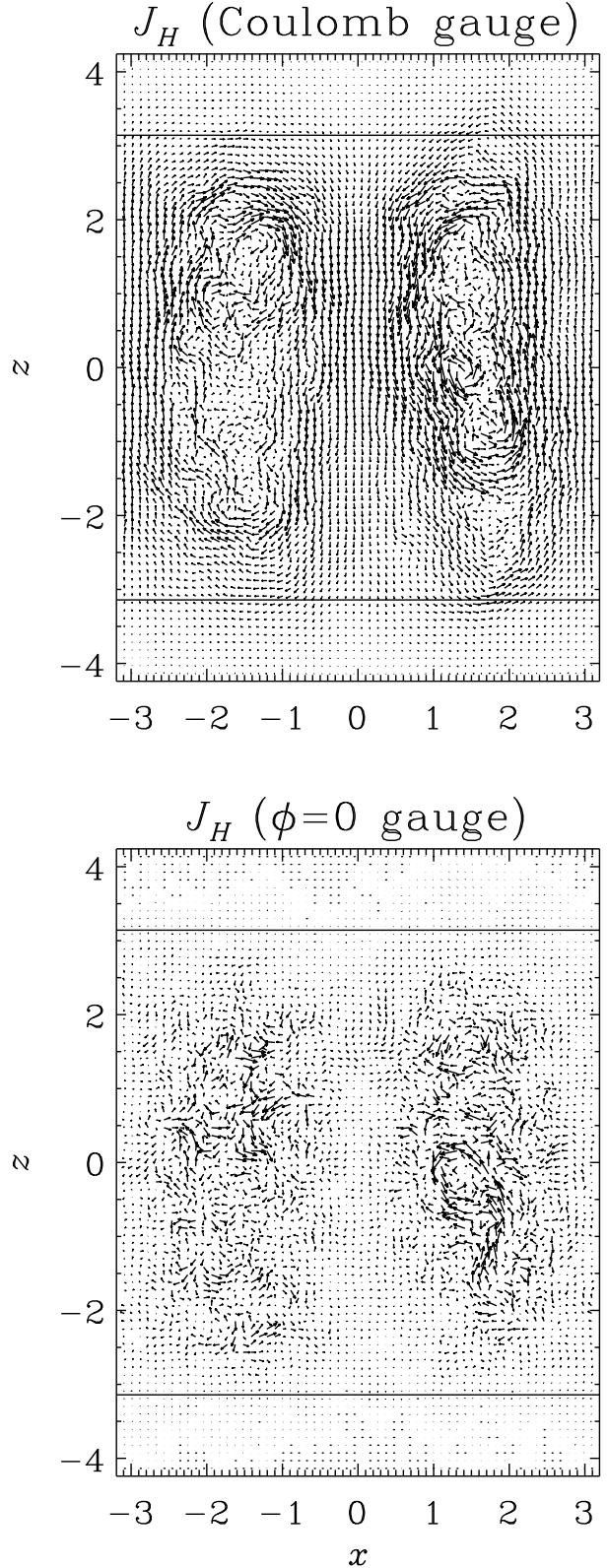}\caption[]{
Magnetic helicities flux density in Coulomb gauge (upper panel)
compared with that in the $\phi=0$ gauge
(lower panel). The local flux density has been averaged in $y$ and
in $t$ (between $\lambda t=32$ and $42$). Run~C.
}\label{Fpethan_average}\end{figure}

\subsection{Magnetic helicity flux}

Vishniac and Cho (2001) pointed out that the $\phi=0$ gauge in \Eq{dAdt}
is not suited for analysing {\it local} helicity flux densities.
We recall that the procedures of Berger \& Field (1984) and BD2001
only apply to helicities in a given volume and the corresponding
helicity fluxes through its bounding surface. Magnetic helicity
densities and the corresponding flux densities are only possible
to define in a given gauge.

In the following we denote by $\AAA_{\rm c}$ the magnetic vector
potential in the Coulomb gauge, whilst $\AAA$ still refers to the
vector potential in the $\phi=0$ gauge. The conversion between the
two is given by $\AAA_{\rm c}=\AAA-\AAA_0$, where $\AAA_0=\bra{\AAA}+\nab\psi$
and $\nabla^2\psi=\nab\cdot\AAA$. In the following we use
the notation $\AAA_0=\nab(\nabla^{-2}\nab\cdot\AAA)$.

In the $\phi=0$ gauge the magnetic helicity flux density is simply
$\EE\times\AAA$, but when converted into the Coulomb gauge it becomes
(see \App{SCoulomb})
\EQ
\JJ_{\rm H}^{\rm Cou}=
(\EE+\EE_0)\times(\AAA-\AAA_0)+\nab\times[2\phi(\AAA-\AAA_0)].
\label{helfludens_Coulomb2}
\EN
The second part does not contribute to the divergence of the
magnetic helicity flux.
In \Fig{Fpethan_average} we plot the first part, denoted by
$\JJ_{\rm H,1st}^{\rm Cou}(\mbox{fluct})$. The suffix `(fluct)'
indicates that we have only included the contribution from the fluctuating
components of $\AAA$ and $\EE$ and then averaged over $y$ and $t$.
This term was used extensively by VC2001. We also compare with the
helicity flux in the $\phi=0$ gauge, $\JJ_{\rm H}^{\phi=0}(\mbox{fluct})$.
It is striking that the two are quite different;
$\JJ_{\rm H,1st}^{\rm Cou}(\mbox{fluct})$ shows a
systematic circulation pattern with noise where shear is weak, i.e.\
near $x=\pm\pi/2$. Such a circulation patter is absent in
$\JJ_{\rm H}^{\phi=0}(\mbox{fluct})$.

\begin{figure}[t!]\epsfbox{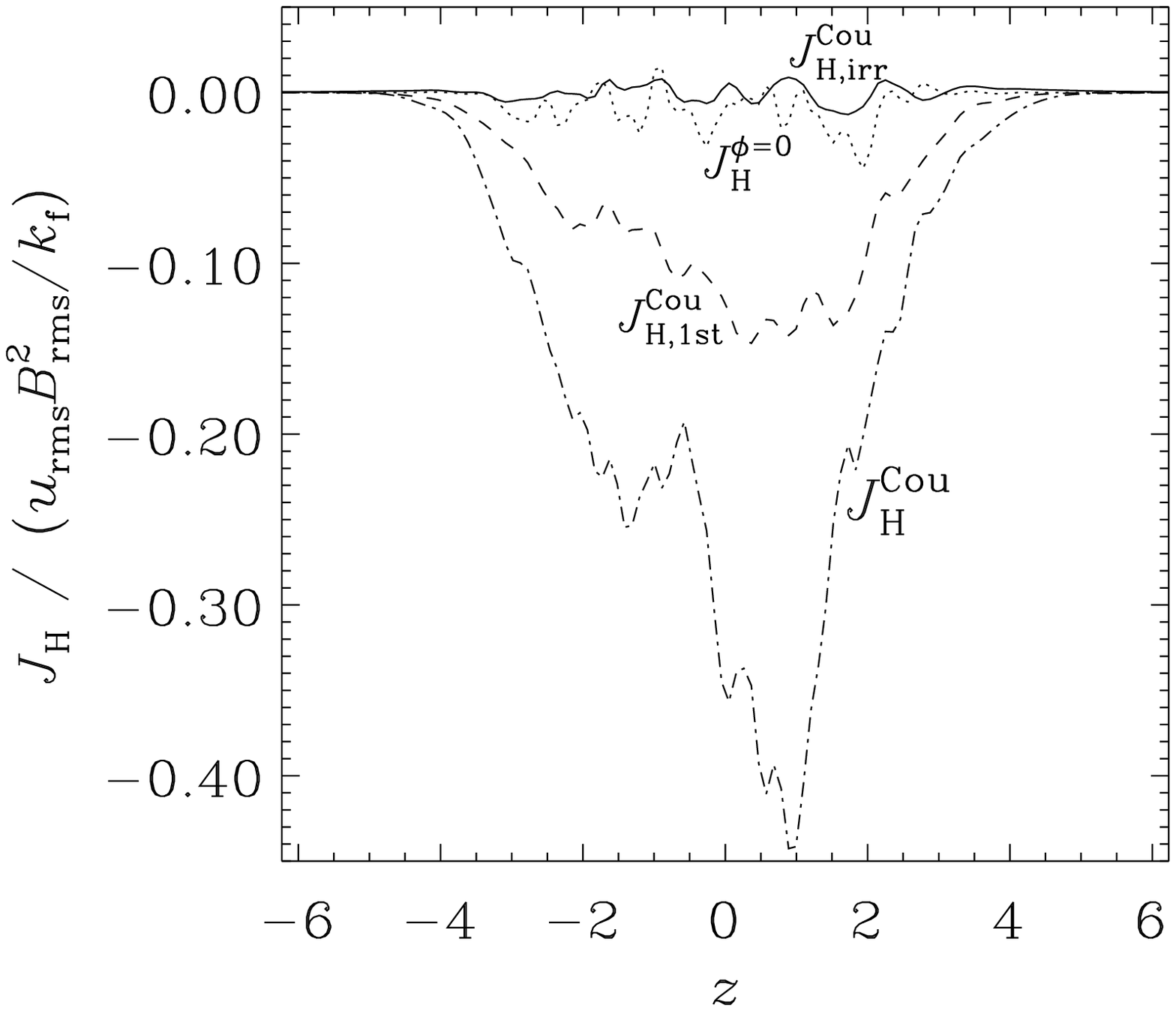}\caption[]{
Time averaged vertical magnetic helicity flux density at $x=0$ as
a function of $z$ in different gauges. $J_{\rm H}^{\phi=0}$ refers
to the $\phi=0$ gauge and $J_{\rm H}^{\rm Cou}$ refers to the
Coulomb gauge. $J_{\rm H,irr}^{\rm Cou}$ refers to the irrotational
part and $J_{\rm H,1st}^{\rm Cou}$ refers to the first part in
\Eq{helfludens_Coulomb2}. Like in \Fig{Fpethan_average}, only the
contributions from the fluctuating components of $\AAA$ and $\EE$
are included. Run~C.
}\label{Fpjzmid}\end{figure}

There may be additional solenoidal contributions also from
$\JJ_{\rm H,1st}^{\rm Cou}$. In general
we may split $\JJ_{\rm H}^{\rm Cou}$ into an irrotational and a
solenoidal part,
\EQ
\JJ_{\rm H}^{\rm Cou}=
\JJ_{\rm H,irr}^{\rm Cou}+\JJ_{\rm H,sol}^{\rm Cou}.
\EN
It turns out that $\JJ_{\rm H,1st}^{\rm Cou}$ has still a
noticeable solenoidal (i.e.\ rotational) contribution.
In particular, it is interesting to note that $\JJ_{\rm H,1st}^{\rm Cou}$
has a systematic flux
through the middle of the domain with an orientation that
agrees with that predicted by VC2001 (upward where $S>0$
and downward where $S<0$).
At $x=0$ (i.e.\ the position where shear is maximum),
the solenoidal part of the averaged magnetic helicity flux is about
200 times larger than its irrotational part. The profiles of the time
averaged vertical magnetic helicity flux density at $x=0$ is plotted in
\Fig{Fpjzmid} as functions of $z$ for different gauges.
Clearly, the relevant irrotational part of the magnetic helicity flux
does not have a systematic component.

\epsfxsize=8.6cm

\begin{figure}[t!]\epsfbox{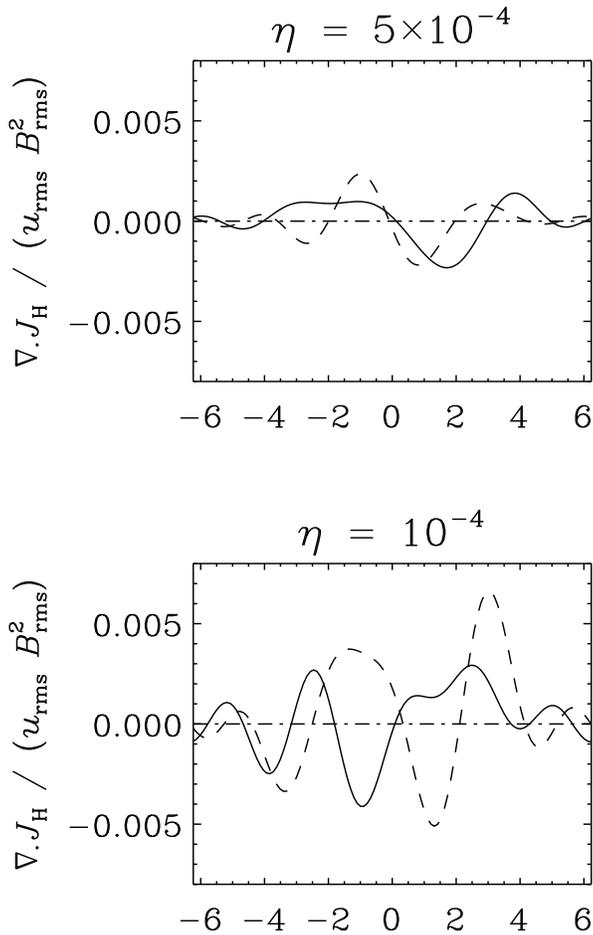}\caption[]{
Divergence of the mean helicity flux, normalized to make it similar
to a local dynamo number. As in \Figs{Fpethan_average}{Fpjzmid}, only
the contributions from the fluctuating components of $\AAA$ and $\EE$
are included in the calculation of the $y$ and $t$ averaged magnetic
helicity flux. Run~C.
}\label{Fpethan_average_div_comp}\end{figure}

In order to assess the possibility of dynamo action we plot in
\Fig{Fpethan_average_div_comp} vertical profiles of the divergence
of the helicity flux density, in units of $u_{\rm rms}B^2_{\rm rms}$.
This non-dimensional quantity is reminiscent of a local dynamo number,
$\alpha_{\rm BH}/(k_1\eta_{\rm T})\approx(k_{\rm f}/k_1)
(\alpha_{\rm BH}/u_{\rm rms})$, where
\EQ
\alpha_{\rm BH}=-\nab\cdot\JJ_{\rm H}/(2\BB^2)
\approx-\nab\cdot\JJ_{\rm H}/B^2_{\rm rms}
\EN
has been introduced for the $\alpha$ effect of Bhattacharjee \& Hameiri (1986) and
$k_{\rm f}\eta_{\rm T}\approx u_{\rm rms}$ for the turbulent magnetic
diffusivity, $\eta_{\rm T}$. Our results
(\Fig{Fpethan_average_div_comp}) suggest that
the local dynamo number is fluctuating in space and that
its rms value is somewhat larger when the magnetic Reynolds number
is larger (cf.\ lower panel).
The fact that $\alpha_{\rm BH}$ fluctuates in space (and probably also
in time) is not necessarily bad; such an `incoherent' $\alpha$ may still, together
with shear and turbulent diffusion, contribute to producing large scale
fields (Vishniac \& Brandenburg 1997).

\FFig{Fpethan_average_div_comp} shows that the non-dimensional measure
of the $\alpha_{\rm BH}$ fluctuates around $\pm0.01$. This value should
be compared with the critical value of $\alpha/(\eta_{\rm T}k_1)\equiv
C_\alpha$ above which dynamo action is possible. We define the dynamo
number as ${\cal D}=C_\alpha C_S$, where $C_S=S/(\eta_{\rm T}k_1^2)$.
With $S\approx S_0=1$ and $\eta_{\rm T}k_1=u_{\rm rms}k_1/k_{\rm f}=0.005$
we have $C_S=200$, and since the critical dynamo number is around
2 (BBS2001) we have 0.01, which agrees with the estimate above
(cf.\ \Fig{Fpethan_average_div_comp}). However, this estimate has been
too optimistic in several ways: the actual value of $S$ is smaller than $S_0$
and the incoherent $\alpha$ effect dynamo will be less efficient.
This may explain why the Vishniac--Cho effect does not seem to operate in
the present simulations, but it may become more
important at higher magnetic Reynolds numbers.

\epsfxsize=8.6cm

\section{Conclusions}
\label{Sconcl}

In this paper we have re-examined the recent suggestion by VC2001 that
large scale dynamo action can result from velocity correlations involving
higher derivatives. We found this effect to be present both in global
accretion disc simulations as well as in models of forced turbulence
with no rotation and just shear. Nevertheless, it
does not seem to produce large scale dynamo action in the parameter
regime considered here. In particular, we find no signs of any net
vertical flux of magnetic helicity through the domain. This was thought
to be an important property of the model suggested by VC2001.  

Of course, the range of parameters considered in the present work
is limited, and the degree of stratification is relatively weak.
Nevertheless, the anticipated velocity correlations are strongly present
and yet there are no signs of large scale dynamo action. Although we
cannot exclude the possibility of different behaviour at larger magnetic
Reynolds number or in more realistic representations of accretion discs,
it is clear that the anticipated effect will not be easily detectable.
This is quite important given the fact that large scale dynamo action
owing to the helicity effect is so much stronger than nonhelical dynamo
action. Thus, if one is
to find the effect anticipated by VC2001 it will be quite important
to isolate it from the much stronger helicity-driven dynamo effect.

Some concluding speculations regarding the viability of helicity-driven
dynamo action in astrophysical settings are now in order.  At large
magnetic Reynolds numbers dynamo action will always generate strong
magnetic fields within a short period of time.  What the helicity
constraint does is to prevent the formation of large scale helical
patterns in a time less than a certain fraction of
the magnetic diffusion time. It
does not exclude however the formation of large scale patterns where the
magnetic helicity cancels to zero. This would however require an exchange
of magnetic helicity between various sub-domains. This does unfortunately
not come automatically, as the simulations of BD2001 have shown.
However, for the sun the relevant resistive time scales are estimated
to be around or less than $10^6$ years.  On the one hand this is long
enough for large scale dynamo to be well established at the present
time. On the other hand, it would suggest that the time scale for the
solar cycle must essentially be controlled by non-resistive effects. One
idea that deserved further attention is the possibility that the dynamo
wave corresponds to actual fluid motions within the solar convection zone,
such that the magnetic helicity within a lagrangian fluid patch remains
to be conserved. Since this would correspond to a systematic flux of
magnetic helicity, this mechanism would be similar to that of VC2001.
Here, however, the magnetic helicity flux would not be self-driven,
but driven externally, e.g.\ by the meridional circulation.
A number of recent investigations have shown that meridional circulation
would be capable of reversing the sense of the dynamo wave driven by the
$\alpha\Omega$-dynamo (Durney 1995, Choudhuri \ea 1995, K\"uker \ea 2001).
This is similar to the possibility discussed above where the dynamo wave itself
drives the meridional circulation.

As far as discs is concerned, the long resistive time scale is perhaps
not a problem, because the possibility of strong outflows always shortens
the saturation time scales, albeit at the expense of lowering the final
saturation field strengths (see BD2001). The final solution to the
problem may require more realistic global simulations with explicit
resistivities, combined with suitable analytic approaches to enable
one to extrapolate to astrophysical conditions.

\begin{acknowledgements}
We thank Wolfgang Dobler, Kandu Subramanian and Ethan Vishniac for interesting
discussions and comments on the manuscript.
Use of the PPARC supported supercomputers in St Andrews and Leicester (UKAFF)
is acknowledged. We thank the John v.\ Neumann-Institut for Computing at the
Forschungszentrum J\"ulich, Germany, for using the T90 computer.
R.A.\ acknowledges the kind support by the Deutsche Forschungsgemeinschaft
and the hospitality of Nordita, where much of this work has been carried out.
\end{acknowledgements}

\appendix

\section{A gauge in which magnetic helicity and magnetic helicity flux are
automatically gauge invariant}
\label{App1}

The purpose of this appendix is to show that there is a particular gauge
for the vector potential of the one-dimensional mean field such that
$\int\overline{\vec{A}}\cdot\overline{\vec{B}}\,\dd z$ is automatically
gauge invariant. This allows us then to define a {\it local} helicity
flux such that its integral over a closed surface (top and bottom of a
slab) equals the gauge invariant integrated helicity flux of BD2001.

The evolution equation of the mean (one-dimensional)
magnetic vector potential is
\EQ
\dot{\overline{\vec{A}}}=-(\overline{\vec{E}}-\overline{\vec{E}}_0),
\label{dAmeandt}
\EN
where $\vec{E}=\eta\mu_0\vec{J}-\vec{u}\times\vec{B}$ is the electric
field, $\overline{\vec{E}}_0=\overline{\vec{E}}_0(t)$ is an integration
constant, and $\overline{\vec{A}}$ and $\overline{\vec{E}}$ depend only on
$z$ and $t$. From \Eq{dAmeandt} follows the evolution equation for
the magnetic helicity density,
\EQ
{\partial\over\partial t}({\overline{\vec{A}}}\cdot{\overline{\vec{B}}})
+\vec{\nabla}\cdot[
(\overline{\vec{E}}+\overline{\vec{E}}_0)\times{\overline{\vec{A}}}]
=-2\overline{\vec{E}}\cdot\overline{\vec{B}}.
\label{dABmeandt}
\EN
In BD2001 the gauge independent magnetic helicity
of the mean field was found to be
\EQ
H_{\rm mean}=\int_{z_1}^{z_2}
\overline{\vec{A}}\cdot\overline{\vec{B}}\,\dd z
+\hat{\vec{z}}\cdot(\overline{\vec{A}}_1\times\overline{\vec{A}}_2),
\label{hel_bd2001}
\EN
where $\overline{\vec{A}}_1$ and $\overline{\vec{A}}_2$ are the values
of $\overline{\vec{A}}$ at $z=z_1$ and $z_2$, respectively. At the initial
time one can always subtract a constant from $\overline{\vec{A}}$ such that
the second term vanishes. This constant turns out to be the average
of $\overline{\vec{A}}_1$ and $\overline{\vec{A}}_2$, so we replace initially
\EQ
\overline{\vec{A}}\rightarrow\overline{\vec{A}}-\overline{\vec{A}}_0,\quad
\mbox{where}\quad
\overline{\vec{A}}_0=\half(\overline{\vec{A}}_1+\overline{\vec{A}}_2).
\EN
Next we choose $\overline{\vec{E}}_0$ such that
$\overline{\vec{A}}_1+\overline{\vec{A}}_2$ remains zero at all later
times. This yields
\EQ
\overline{\vec{E}}_0=\half(\overline{\vec{E}}_1+\overline{\vec{E}}_2).
\EN
We can then express the two integrated fluxes on $z_2$ as
\EQ
Q_{\rm mean}^{(2)}=\hat{\vec{z}}\cdot[
(\overline{\vec{E}}_2+\overline{\vec{E}}_0)\times{\overline{\vec{A}}}_2]
=\hat{\vec{z}}\cdot[({\textstyle{3\over2}}\overline{\vec{E}}_2
 +{\textstyle{1\over2}}\overline{\vec{E}}_1)\times{\overline{\vec{A}}}_2].
\EN
Using the fact that $\overline{\vec{A}}_1+\overline{\vec{A}}_2=0$ we
can write
\EQ
{\textstyle\int_{z_1}^{z_2}}\overline{\vec{B}}\,\dd z
=\hat{\vec{z}}\times(\overline{\vec{A}}_2-\overline{\vec{A}}_1)
=2\hat{\vec{z}}\times\overline{\vec{A}}_2
=-2\hat{\vec{z}}\times\overline{\vec{A}}_1.
\EN
This allows us to express $Q_{\rm mean}^{(2)}$ as
\EQ
Q_{\rm mean}^{(2)}=-({\textstyle{3\over4}}\overline{\vec{E}}_2
     +{\textstyle{1\over4}}\overline{\vec{E}}_1)
\cdot{\textstyle\int_{z_1}^{z_2}}\overline{\vec{B}}\,\dd z.
\EN
At $z=z_1$ we count the flux as negative when helicity leaves the domain
in the downward direction, so
\EQ
Q_{\rm mean}^{(1)}=+({\textstyle{3\over4}}\overline{\vec{E}}_1
     +{\textstyle{1\over4}}\overline{\vec{E}}_2)
\cdot{\textstyle\int_{z_1}^{z_2}}\overline{\vec{B}}\,\dd z,
\EN
with $Q_{\rm mean}=Q_{\rm mean}^{(2)}-Q_{\rm mean}^{(1)}$ being the
gauge invariant magnetic helicity flux of BD2001.
The average upward flux of mean magnetic helicity on the two boundaries
is
\EQ
Q_{\rm mean}^{\rm(up)}
\!=\!\half\!\left[Q_{\rm mean}^{(2)}\!+\!Q_{\rm mean}^{(1)}\right]
\!=\!-\quarter(\overline{\vec{E}}_2-\overline{\vec{E}}_1)
\cdot{\textstyle\int_{z_1}^{z_2}}\overline{\vec{B}}\,\dd z.\;
\EN

\section{Magnetic helicity in sub-domains}
\label{App2}

In a periodic domain the sum of the magnetic helicities of
any two sub-domains is equal
to the magnetic helicity of the entire (periodic) domain.
In the non-periodic case this is not the case if the gauge
invariant magnetic helicity of BD2001 is used also for the sub-domains.
We therefore calculate the magnetic helicity of sub-domains by using
the gauge discussed in \App{App1}, so the magnetic helicity between
the points $z_{\it a}$ and $z_{\it b}$ is then
\EQ
H_{\rm mean}^{\it(ab)}=\int_{z_{\it a}}^{z_{\it b}}
(\overline{\vec{A}}-\overline{\vec{A}}_0)
\cdot\overline{\vec{B}}\,\dd z,
\label{maghel_subdomain}
\EN
where $\overline{\vec{A}}_0=\half(\overline{\vec{A}}_1+
\overline{\vec{A}}_2)$ is independent of the values of
$z_{\it a}$ and $z_{\it b}$. For $z_{\it a}=z_1$ and $z_{\it b}=z_2$
we recover \Eq{hel_bd2001}, and the sum of the magnetic
helicities of sub-domains agrees with the magnetic helicity of
the whole domain from $z_1$ to $z_2$.

Similar to \Eq{dABmeandt}, we can now derive an evolution equation for
$(\overline{\vec{A}}-\overline{\vec{A}}_0)\cdot\overline{\vec{B}}$. The
flux term is then like in \Eq{dABmeandt}, but with $\overline{\vec{A}}$
being replaced by $\overline{\vec{A}}-\overline{\vec{A}}_0$.
The magnetic helicity flux out of an individual sub-domain is then
$Q_{\rm mean}^{\it(ab)} =Q_{\rm mean}^{\it(b)}-Q_{\rm mean}^{\it(a)}$, where
\EQ
Q_{\rm mean}^{(\alpha)}=\hat{\vec{z}}\cdot[
(\overline{\vec{E}}_{\alpha}+\overline{\vec{E}}_0)
\times({\overline{\vec{A}}}_{\alpha}-{\overline{\vec{A}}}_0)],
\quad\alpha=a,b.
\label{maghelflux_subdomain}
\EN
Again, the sum of net helicity fluxes out of sub-domains equals the
gauge invariant net helicity fluxes, $Q_{\rm mean}$, of the full domain.
This formula can be applied to value of $z_\alpha$, in particular to
the equator. In that case one obtains the horizontally averaged
magnetic helicity flux through the surface $z_\alpha=0$.

\section{Velocity gradient correlation tensor}
\label{App3}

In order to check that the Vishniac--Cho correlation is the dominant
correlation among the different components of the velocity gradient
matrix, we define the correlation coefficient
\EQ
C_{ijkl}={\bra{u_{i,j}u_{k,l}}\over\sqrt{\bra{u_{i,j}^2}\bra{u_{k,l}^2}}}.
\EN
We denote by $C^{(\pm)}_{ijkl}$ the values of $C_{ijkl}$ evaluated in
those sub-domains where the sign of shear is locally positive or negative,
respectively. We separate the coefficients that are symmetric and
antisymmetric with respect to changing the sign of shear. Hence, we calculate
\EQ
C^{(S)}_{ijkl}=\half[C^{(+)}_{ijkl}+C^{(-)}_{ijkl}],\quad
C^{(A)}_{ijkl}=\half[C^{(+)}_{ijkl}-C^{(-)}_{ijkl}],
\EN
whose values are shown in \Tabs{Tsym}{Tantisym}, respectively.

\begin{table}[t!]\caption{
The coefficients $C^{(S)}_{ijkl}$, arranged in blocks where $k$
increases downward and $l$ increases to the right. Within each block
$i$ increases downward and $j$ increases to the right. In bold are
given the values that are largest by magnitude, but different from
unity.\label{Tsym}}\begin{small}\begin{tabular}{rrr|rrr|rrr}
    &  $\!l=1\!$   &      &       &  $\!l=2\!$  &      &     &  $\!l=3\!$   &  $\!\!\!$\\
\hline
$\!\!\!\!${    1.00}&$\!\!\!\!${    0.04}&$\!\!\!\!${   -0.09} &$\!\!\!\!${    0.04}&$\!\!\!\!${    1.00}&$\!\!\!\!${   -0.14} &$\!\!\!\!${   -0.09}&$\!\!\!\!${   -0.14}&$\!\!\!\!${    1.00} $\!\!\!$\\
$\!\!\!\!${    0.02}&$\!\!\!\!${   -0.01}&$\!\!\!\!${   -0.01} &$\!\!\!\!${\bf-0.26}&$\!\!\!\!${   -0.03}&$\!\!\!\!${    0.03} &$\!\!\!\!${   -0.03}&$\!\!\!\!${    0.01}&$\!\!\!\!${    0.03} $\!\!\!$\\
$\!\!\!\!${    0.01}&$\!\!\!\!${    0.05}&$\!\!\!\!${   -0.23} &$\!\!\!\!${    0.10}&$\!\!\!\!${    0.10}&$\!\!\!\!${   -0.02} &$\!\!\!\!${\bf-0.54}&$\!\!\!\!${   -0.01}&$\!\!\!\!${    0.04} $\!\!\!$\\
\hline
$\!\!\!\!${    0.02}&$\!\!\!\!${\bf-0.26}&$\!\!\!\!${   -0.03} &$\!\!\!\!${   -0.01}&$\!\!\!\!${   -0.03}&$\!\!\!\!${    0.01} &$\!\!\!\!${   -0.01}&$\!\!\!\!${    0.03}&$\!\!\!\!${    0.03} $\!\!\!$\\
$\!\!\!\!${    1.00}&$\!\!\!\!${   -0.00}&$\!\!\!\!${   -0.03} &$\!\!\!\!${   -0.00}&$\!\!\!\!${    1.00}&$\!\!\!\!${   -0.01} &$\!\!\!\!${   -0.03}&$\!\!\!\!${   -0.01}&$\!\!\!\!${    1.00} $\!\!\!$\\
$\!\!\!\!${   -0.03}&$\!\!\!\!${    0.02}&$\!\!\!\!${   -0.04} &$\!\!\!\!${    0.00}&$\!\!\!\!${    0.07}&$\!\!\!\!${   -0.07} &$\!\!\!\!${   -0.03}&$\!\!\!\!${\bf-0.60}&$\!\!\!\!${    0.06} $\!\!\!$\\
\hline
$\!\!\!\!${    0.01}&$\!\!\!\!${    0.10}&$\!\!\!\!${\bf-0.54} &$\!\!\!\!${    0.05}&$\!\!\!\!${    0.10}&$\!\!\!\!${   -0.01} &$\!\!\!\!${   -0.23}&$\!\!\!\!${   -0.02}&$\!\!\!\!${    0.04} $\!\!\!$\\
$\!\!\!\!${   -0.03}&$\!\!\!\!${    0.00}&$\!\!\!\!${   -0.03} &$\!\!\!\!${    0.02}&$\!\!\!\!${    0.07}&$\!\!\!\!${\bf-0.60} &$\!\!\!\!${   -0.04}&$\!\!\!\!${   -0.07}&$\!\!\!\!${    0.06} $\!\!\!$\\
$\!\!\!\!${    1.00}&$\!\!\!\!${   -0.00}&$\!\!\!\!${   -0.03} &$\!\!\!\!${   -0.00}&$\!\!\!\!${    1.00}&$\!\!\!\!${   -0.13} &$\!\!\!\!${   -0.03}&$\!\!\!\!${   -0.13}&$\!\!\!\!${    1.00} $\!\!\!$\\
\hline\end{tabular}\end{small}\end{table}

\begin{table}[t!]\caption{
Like \Tab{Tsym}, but for the coefficients $C^{(A)}_{ijkl}$ which have
an antisymmetric dependence on shear. The components that enter the
Vishniac--Cho correlation are shown in bold;
these are also the coefficients with the largest magnitude in
this table. \label{Tantisym}
}\begin{small}\begin{tabular}{rrr|rrr|rrr}
    &  $\!\!l=1\!\!$   &      &       &  $\!\!l=2\!\!$  &      &     &  $\!\!l=3\!\!$   &  $\!\!\!$\\
\hline
$\!\!\!\!${    0.00}&$\!\!\!\!${   -0.16}&$\!\!\!\!${   -0.01} &$\!\!\!\!${   -0.16}&$\!\!\!\!${    0.00}&$\!\!\!\!${   -0.06} &$\!\!\!\!${   -0.01}&$\!\!\!\!${   -0.06}&$\!\!\!\!${    0.00} $\!\!\!$\\
$\!\!\!\!${   -0.01}&$\!\!\!\!${    0.00}&$\!\!\!\!${    0.00} &$\!\!\!\!${    0.01}&$\!\!\!\!${   -0.17}&$\!\!\!\!${    0.09} &$\!\!\!\!${    0.00}&$\!\!\!\!${    0.02}&$\!\!\!\!${   -0.13} $\!\!\!$\\
$\!\!\!\!${   -0.15}&$\!\!\!\!${   -0.02}&$\!\!\!\!${    0.00} &$\!\!\!\!${   -0.03}&$\!\!\!\!${   -0.04}&$\!\!\!\!${    0.21} &$\!\!\!\!${    0.01}&$\!\!\!\!${\bf 0.28}&$\!\!\!\!${   -0.07} $\!\!\!$\\
\hline
$\!\!\!\!${   -0.01}&$\!\!\!\!${    0.01}&$\!\!\!\!${    0.00} &$\!\!\!\!${    0.00}&$\!\!\!\!${   -0.17}&$\!\!\!\!${    0.02} &$\!\!\!\!${    0.00}&$\!\!\!\!${    0.09}&$\!\!\!\!${   -0.13} $\!\!\!$\\
$\!\!\!\!${    0.00}&$\!\!\!\!${   -0.04}&$\!\!\!\!${    0.11} &$\!\!\!\!${   -0.04}&$\!\!\!\!${    0.00}&$\!\!\!\!${   -0.01} &$\!\!\!\!${    0.11}&$\!\!\!\!${   -0.01}&$\!\!\!\!${    0.00} $\!\!\!$\\
$\!\!\!\!${    0.01}&$\!\!\!\!${    0.03}&$\!\!\!\!${    0.10} &$\!\!\!\!${    0.00}&$\!\!\!\!${    0.03}&$\!\!\!\!${   -0.00} &$\!\!\!\!${    0.05}&$\!\!\!\!${   -0.02}&$\!\!\!\!${    0.01} $\!\!\!$\\
\hline
$\!\!\!\!${   -0.15}&$\!\!\!\!${   -0.03}&$\!\!\!\!${    0.01} &$\!\!\!\!${   -0.02}&$\!\!\!\!${   -0.04}&$\!\!\!\!${\bf 0.28} &$\!\!\!\!${    0.00}&$\!\!\!\!${    0.21}&$\!\!\!\!${   -0.07} $\!\!\!$\\
$\!\!\!\!${    0.01}&$\!\!\!\!${    0.00}&$\!\!\!\!${    0.05} &$\!\!\!\!${    0.03}&$\!\!\!\!${    0.03}&$\!\!\!\!${   -0.02} &$\!\!\!\!${    0.10}&$\!\!\!\!${   -0.00}&$\!\!\!\!${    0.01} $\!\!\!$\\
$\!\!\!\!${    0.00}&$\!\!\!\!${\bf-0.51}&$\!\!\!\!${    0.10} &$\!\!\!\!${\bf-0.51}&$\!\!\!\!${    0.00}&$\!\!\!\!${    0.00} &$\!\!\!\!${    0.10}&$\!\!\!\!${    0.00}&$\!\!\!\!${    0.00} $\!\!\!$\\
\hline\end{tabular}\end{small}\end{table}

In \Tab{Tsym} the largest contributions come from $C^{(S)}_{xyyx}=-0.26$,
$C^{(S)}_{yzzy}=-0.60$, and $C^{(S)}_{zxxz}=-0.54$. In \Tab{Tantisym}
the largest contributions come from $C^{(A)}_{xzzy}=+0.28$ and
$C^{(A)}_{zxzy}=-0.51$.
These are also the coefficients that are important in the Vishniac--Cho correlation.

\section{Coulomb-gauged helicity flux}
\label{SCoulomb}

In the $\phi=0$ gauge we have $\dot{\AAA}=-\EE$. In that gauge,
the evolution of the magnetic helicity density is given by
\EQ
{\partial\over\partial t}(\AAA\cdot\BB)+\nab\cdot(\EE\times\AAA)
=-2\EE\cdot\BB.
\label{heldens_nophi}
\EN
Note that the term on the right hand side of this equation is
gauge-invariant, but the terms on the left hand side are not.
In the Coulomb gauge we have $\nab\cdot\AAA_{\rm C}=0$, where
$\AAA_{\rm C}\equiv\AAA-\AAA_0$. In order to maintain
$\nab\cdot\AAA_{\rm C}=0$ for all times, we have to add the term
$-\nab\phi\equiv\EE_0=\nab(\nabla^{-2}\nab\cdot\EE)$ to the
right hand side of the uncurled induction equation,
\EQ
{\partial\over\partial t}\AAA_{\rm C}=-(\EE-\EE_0).
\EN
In this gauge the evolution equation for the helicity density becomes
\EQ
{\partial\over\partial t}(\AAA_{\rm C}\cdot\BB)
+\nab\cdot[(\EE-\EE_0)\times\AAA_{\rm C}]
=-2(\EE-\EE_0)\cdot\BB.
\label{heldens_Coulomb}
\EN
Unlike \Eq{heldens_nophi}, the right hand side of \Eq{heldens_Coulomb}
is gauge dependent
owing to the extra term $2\EE_0\cdot\BB$. However, because of $\nab\cdot\BB=0$,
we can write this as the divergence of another contribution to the
helicity flux density,
\EQ
2\EE_0\cdot\BB=-2(\nab\phi)\cdot\BB=-\nab\cdot(2\phi\BB),
\EN
which should be included in the expression for the Coulomb gauged
helicity flux density,
\EQ
\JJ_{\rm H}^{\rm Cou}=(\EE-\EE_0)\times(\AAA-\AAA_0)+2\phi\BB,
\label{helfludens_Coulomb1}
\EN
so \Eq{heldens_Coulomb} becomes
\EQ
{\partial\over\partial t}(\AAA_{\rm C}\cdot\BB)
+\nab\cdot\JJ_{\rm H}^{\rm Cou}=-2\EE\cdot\BB.
\label{heldens_Coulomb2}
\EN
Now the right hand sides of \Eqs{heldens_nophi}{heldens_Coulomb2}
agree and are gauge-invariant. Note, however, that
\EQ
\phi\BB=\phi\nab\times\AAA_{\rm C}
=\nab\times(\phi\AAA_{\rm C})+\EE_0\times\AAA_{\rm C},
\EN
so \Eq{helfludens_Coulomb1} can also be written as
\EQ
\JJ_{\rm H}^{\rm Cou}=
(\EE+\EE_0)\times(\AAA-\AAA_0)+\nab\times[2\phi(\AAA-\AAA_0)],
\EN
which is identical to \Eq{helfludens_Coulomb2}.


\begin{thebibliography}{99}

\bibitem[]{}
Arlt R., R\"udiger G.\pana{2001}
{\sf astro-ph/0101470}

\bibitem[]{}
Armitage P. J.\yapj{1998}{501}{L189}

\bibitem[]{}
Berger M., Field G. B.\yjfm{1984}{147}{133}

\bibitem[]{}
Bhattacharjee A., Hameiri E.\yprl{1986}{57}{206}

\bibitem[]{}
Bhattacharjee A., Yuan Y.\yapj{1995}{449}{739}

\bibitem[]{}
Blackman E. G., Field G. F.\yapj{2000}{534}{984}

\bibitem[]{}
Brandenburg A.\yproc{1998}{61}
{Theory of Black Hole Accretion Discs}
{M. A. Abramowicz, G. Bj\"ornsson \& J. E. Pringle}
{Cambridge University Press}

\bibitem[]{}
Brandenburg A.\yapj{2001}{550}{824}

\bibitem[]{}
Brandenburg A., Dobler W.\yana{2001}{369}{329}

\bibitem[]{}
Brandenburg A., Bigazzi A., Subramanian K.\pmn{2001}
{\sf astro-ph/0011081} (BBS2001)

\bibitem[]{}
Brandenburg A., Nordlund \AA., Stein R. F.,
Torkelsson U.\yapj{1995}{446}{741}

\bibitem[]{}
Brandenburg A., Jennings R. L., Nordlund \AA.,
Rieutord M., Stein R. F., Tuominen I.\yjfm{1996}{306}{325}

\bibitem[]{}
Cattaneo F., Hughes D. W.\ypr{1996}{E 54}{R4532}

\bibitem[]{}
Cattaneo F., Vainshtein S. I.\yapjl{1991}{376}{L21}

\bibitem[]{}
Choudhuri A. R., Sch\"ussler M., Dikpati M.\yana{1995}{303}{L29}

\bibitem[]{}
Durney B. R.\ysph{1995}{166}{231}

\bibitem[]{}
Frisch U., Pouquet A., L\'eorat J., Mazure A.\yjfm{1975}{68}{769}

\bibitem[]{}
Gilbert A. D., Frisch U., Pouquet A.\ygafd{1988}{42}{151}

\bibitem[]{}
Glatzmaier G. A., Roberts P. H.\ynat{1995}{377}{203}

\bibitem[]{}
Gruzinov A. V., Diamond P. H.\ypp{1995}{2}{1941}

\bibitem[]{}
Hawley J. F.\yapj{2000}{528}{462}

\bibitem[]{}
Krause F., R\"adler K.-H.\ybook{1980}
{Mean-Field Magnetohy\-dro\-dynamics and Dynamo Theory}
{Pergamon Press, Oxford}

\bibitem[]{}
K\"uker M., R\"udiger G., Schultz M.\pana{2001}

\bibitem[]{}
Moffatt H. K.\ybook{1978}
{Magnetic Field Generation in Electrically Conducting Fluids}
{CUP, Cambridge}

\bibitem[]{}
Pouquet A., Frisch U., L\'eorat J.\yjfm{1976}{77}{321}

\bibitem[]{}
R\"udiger G., Pipin V. V.\yana{2000}{362}{756}

\bibitem[]{}
Stone J. M., Norman M.\yapj{1992a}{80}{753}

\bibitem[]{}
Stone J. M., Norman M.\yapj{1992b}{80}{791}

\bibitem[]{}
Stone J. M., Mihalas D., Norman M.\yapj{1992}{80}{791}

\bibitem[]{}
Vainshtein S. I., Cattaneo F.\yapj{1992}{393}{165}

\bibitem[]{}
Vishniac E. T., Brandenburg A.\yapj{1997}{475}{263}

\bibitem[]{}
Vishniac E. T., Cho J.\yapj{2001}{550}{752}

\bibitem[]{}
Zheligovsky V. A., Podvigina O. M., Frisch, U.\pgafd{2001}
{\sf nlin.CD/0012005}

\bibitem[]{}
Ziegler U., R\"udiger G.\yana{2000}{356}{1141}

\end{thebibliography}
\end{document}